\documentclass[10pt,conference]{IEEEtran}
\usepackage{cite}
\usepackage{amsmath,amssymb,amsfonts}
\usepackage{algorithmic}
\usepackage{graphicx}
\usepackage{textcomp}
\usepackage{xcolor}
\usepackage[hyphens]{url}
\usepackage{fancyhdr}
\usepackage[colorlinks=true,citecolor=blue,linkcolor=blue,urlcolor=black,]{hyperref}

\usepackage{enumitem}
\usepackage{caption}

\usepackage{soul}

\usepackage[symbol]{footmisc}

\usepackage[linesnumbered,ruled,vlined]{algorithm2e}
\SetKwInput{KwInput}{Input}                % Set the Input
\SetKwInput{KwOutput}{Output}              % set the Output
\newtheorem{property}{Property}[section]

\newcommand{\pn}{{Tessel}}%\textsc

\newcommand{\eg}{{\it e.g.,\ }}

\newcommand{\ie}{{\it i.e.,\ }}

\definecolor{dkgreen}{rgb}{0,0.6,0}
\definecolor{gray}{rgb}{0.5,0.5,0.5}
\definecolor{mauve}{rgb}{0.58,0,0.82}
\definecolor{olivegreen}{rgb}{0, 0.6, 0}
\definecolor{skyblue}{rgb}{0, 0.4, 1}
\definecolor{antiquefuchsia}{rgb}{0.57, 0.36, 0.51}
\definecolor{fuchsia}{rgb}{1.0, 0.0, 1.0}
\definecolor{darkgoldenrod}{rgb}{0.72, 0.53, 0.04}
\definecolor{glaucous}{rgb}{0.38, 0.51, 0.71}
\definecolor{hanblue}{rgb}{0.27, 0.42, 0.81}
\definecolor{lava}{rgb}{0.81, 0.06, 0.13}

\newcommand{\para}[1]{{\bf \noindent #1 \hspace{2pt}}}

\newcommand{\hpcayear}{2024}

\title{\pn: Boosting Distributed Execution of Large DNN Models via Flexible Schedule Search}
\def\hpcacameraready{} % Uncomment to build camera-ready version

\newcommand\hpcaauthors{
Zhiqi Lin$^{\dagger\ast}$,
Youshan Miao$^{\ddagger}$,
Guanbin Xu$^\dagger$,
Cheng Li$^{\dagger}$,
Olli Saarikivi$^\ddagger$,
Saeed Maleki$^\ddagger$,
Fan Yang$^\ddagger$
}
\newcommand\hpcaaffiliation{
$^\dagger$University of Science and Technology of China,
$^\ddagger$Microsoft Research
}
\newcommand\hpcaemail{
zhiqi.0@mail.ustc.edu.cn,
yomia@microsoft.com,
xugb@mail.ustc.edu.cn,
chengli7@ustc.edu.cn, \\
olli.saarikivi@microsoft.com,
saemal@microsoft.com,
fanyang@microsoft.com
}

\author{
  \ifdefined\hpcacameraready
    \IEEEauthorblockN{\hpcaauthors{}}
      \IEEEauthorblockA{
        \hpcaaffiliation{} \\
        \hpcaemail{}
      }
  \else
    \IEEEauthorblockN{\normalsize{HPCA \hpcayear{} Submission
      \textbf{\#\hpcasubmissionnumber{}}} \\
      \IEEEauthorblockA{
        Confidential Draft \\
        Do NOT Distribute!!
      }
    }
  \fi 
}

\fancypagestyle{camerareadyfirstpage}{%
  \fancyhead{}
  
  \fancyhead[C]{
    \ifdefined\aeopen
    \parbox[][12mm][t]{13.5cm}{}    
    \else
      \ifdefined\aereviewed
      \parbox[][12mm][t]{13.5cm}{}
      \else
      \ifdefined\aereproduced
      \parbox[][12mm][t]{13.5cm}{}
      \else
      \parbox[][0mm][t]{13.5cm}{}
    \fi 
    \fi 
    \fi 
    \ifdefined\aeopen 
      \includegraphics[width=12mm,height=12mm]{ae-badges/open-research-objects.pdf}
    \fi 
    \ifdefined\aereviewed
      \includegraphics[width=12mm,height=12mm]{ae-badges/research-objects-reviewed.pdf}
    \fi 
    \ifdefined\aereproduced
      \includegraphics[width=12mm,height=12mm]{ae-badges/results-reproduced.pdf}
    \fi
  }
  \fancyfoot[C]{}
}
\begin{document}
\maketitle

%Enables the camera ready header and footer
\ifdefined\hpcacameraready 
  \thispagestyle{camerareadyfirstpage}
  \pagestyle{empty}
\else
  \thispagestyle{plain}
  \pagestyle{plain}
\fi

\newcommand{\hpcaheight}{0mm}
\ifdefined\eaopen
\renewcommand{\hpcaheight}{12mm}
\fi

\footnotetext{
$^\ast$This work was done when the author was with Microsoft Research.
}

%%%%%%%%%%%%%%%%%%%%%%%%%%%%%%%%%%%%%%%%
%%%%%%%% -- PAPER CONTENT STARTS -- %%%%%%%%%

\begin{abstract}

Increasingly complex and diverse deep neural network (DNN) models necessitate distributing the execution across multiple devices for training and inference tasks, % along with 
and also require
carefully planned schedules for performance. 
However, existing practices often rely on predefined schedules that may not fully exploit the benefits of emerging diverse model-aware operator placement strategies. Handcrafting high-efficiency schedules
can be challenging due to the large and varying schedule space. This paper presents \pn{}, an automated system that searches for efficient schedules for distributed DNN training and inference for diverse operator placement strategies.
To reduce search costs, \pn{} leverages the insight that the most efficient schedules often exhibit repetitive pattern (\textit{repetend}) across different data inputs. This leads to a two-phase approach: repetend construction and schedule completion. By exploring schedules for various operator placement strategies, \pn{} significantly improves both training and inference performance. Experiments with representative DNN models demonstrate that \pn{} achieves up to 5.5$\times$ training performance speedup and up to 38\% inference latency reduction.

\end{abstract}

\section{Introduction}

% 1. DNN models are getting larger and diverse. Distributed DNN model training and inference are important.
\noindent Deep Neural Network (DNN) models have demonstrated impressive performance across a wide range of domains~\cite{gpt-4, swin-v2, flava, mt5}. As their complexity and depth continue to increase, their size has outpaced the capacity of existing hardware to keep up with their training and inference demands~\cite{gpt-3, swin-v2}. Consequently, the distributed execution of large DNN models across multiple devices has emerged as a necessary solution to mitigate memory limitations. Specifically, pipeline parallelism~\cite{gpipe, dapple, chimera} becomes one of the most adopted techniques 
for efficient parallel execution of distributed DNN training and inference~\cite{megatron3, alpa, deepspeed, varuna}.
% to distribute DNN training and inference tasks among devices for efficient parallel execution.

% 2. a training/inference iteration can be defined as an execution plan
A DNN model consists of computational operators and data tensors, with a training or inferring task iteratively computing the operators over tensors.
The execution of each iteration can be defined by a distributed \emph{execution plan}, which determines spatial placement and temporal schedule of operators among devices. The spatial placement decides whether operators are executed on one or multiple devices, following the temporal schedule that decides per-device execution order of operators.

% 3. temporal schedule is critical to performance, but has not been well explored
Both the spatial placement and temporal schedule are critical to the performance of an execution plan. Prior research has extensively explored various spatial placement strategies and demonstrated their effectiveness in practical applications~\cite{alpa, piper, megatron3, superscaler}. However, existing practices primarily rely on pre-defined schedules~\cite{dapple, chimera, varuna}, which lag behind the advancements in placement strategies and can lead to inefficiencies~(\S\ref{sec:motivation}).

% 4. the problem that this paper aims to solve and the challenges:
% 1) the space is large; 2) space varies due to operator placement strategies
In this paper, we focus on exploring the temporal schedule for distributed DNN execution under diverse spatial placement strategies. Once the operators' placement is determined, the task of finding efficient temporal schedules becomes challenging and complex.
%Exploring efficient spatial placement strategies has been well studied and shows effectiveness in real practice~\cite{alpa, piper, megatron3}. However, upon determining the placement for operators, finding efficient temporal schedules is challenging and complex.
%
First, the schedule space is large. 
% A training iteration usually contains hundreds or even thousands of independent \emph{micro-batches}~\cite{gpipe, megatron2}, 
To mitigate peak memory costs, a training iteration is usually used to divide numerous input data samples into hundreds or even thousands of independent \emph{micro-batches}~\cite{gpipe, megatron2},
thereby creating a large schedule space during execution. For example, one can execute the micro-batches sequentially, 
resulting in minimal peak memory but low device utilization due to idle wait time~\cite{gpipe}.
% which can lead to low device utilization due to idle wait time. 
Alternatively, one can explore more complex schemes by 
% each device executes different 
allowing multiple in-flight % on-the-fly 
micro-batches simultaneously~\cite{dapple, pipedream}, leading to better utilization but requiring careful design of sophisticated schedules 
that can be error-prone.
%to ensure correctness and prevent deadlocks.
%
Furthermore, the schedule space can vary depending on the operator placement. For example, a placement scheme may group model operators into consecutive stages,
% \TODO{execution block in background section?},
with each stage placed to a distinct device~\cite{pipedream}. In such a case, each device is tasked with scheduling the execution of only one stage with different micro-batches. However, an alternative scheme may place multiple stages on the same device~\cite{megatron2, chimera}, which can substantially increase the schedule space, since each device may execute multiple stages following different orders.

Faced with the complexity of the schedule problem, we present \pn{}, a system that takes an operator placement strategy as the input and automatically searches for highly efficient schedules for distributed DNN training and inference. 
We observed that, given the same operator placement, the schedule space for fewer micro-batches is much smaller than for a larger number of micro-batches.
After analyzing a large number of efficient schedules,
% we observe that mostly the schedules repeatedly perform certain execution patterns,
we further observed that efficient schedules often exhibit a repetitive execution pattern,
referred to as \emph{repetend}~(\S\ref{sec:prob-insight}),
where only a small portion of time is dedicated to performing non-repetend executions 
% in the beginning and the end.
at the beginning and the end.
% And the repetend phase usually takes the majority of the execution time, with a small warmup phase in the beginning and a cooldown phase in the end.
The schedule search space can be significantly reduced 
% if we search a small number of micro-batches for such a repetend, and then extend the repetend to construct the schedule for more micro-batches.
by first searching for a repetend within a small number of micro-batches and then extending it to construct the schedule for further micro-batches.
Based on this observation, we introduced a two-phase approach for \pn{} to address the schedule problem, consisting of repetend construction and schedule completion. 
% \pn{} presents a schedule with three parts: warmup, repetend and cooldown. 
During repetend construction, \pn{} searches for an operator set that can 
% \TODO{feasible op set=schedule?} 
form an efficient repetend over a small number of micro-batches~(\S\ref{sec:design-construct}).
% that can be exactly combined to a single one \TODO{one means repetend?}, with the aim of minimizing device idle time resulting from data dependency waiting~(\S\ref{sec:design-construct}).
In the second phase, \pn{} completes the schedule of warmup and cooldown parts by optimizing the execution time for the remaining micro-batch computations~(\S\ref{sec:design-completion}).
% , to further minimize the overall execution time~(\S\ref{sec:design-completion}).
% Finally, the searched schedule is instantiated to the runtime by repeating the repetend component until the desired number of micro-batch is reached. Asynchronous communication is applied to the runtime plan to further improve the performance~(\S\ref{sec:design-inst}).
Finally, the searched schedule is instantiated to the runtime and optimized by inserting non-blocking communication primitives~(\S\ref{sec:design-inst}).

% 8. our results
The automated search in \pn{} facilitates the exploration of schedules for various operator placement strategies and generates highly efficient schedules, including existing ones such as 1F1B~\cite{dapple} or Chimera~\cite{chimera}, as well as novel schedules that can significantly improve training and inference performance. Our experiments demonstrate that \pn{} achieves up to 5.5$\times$ performance speedup on training language models with large embedding layers, and reduces up to 38\% inference latency in multi-modality models such as Flava~\cite{flava}~(\S\ref{sec:eval}). We plan to release \pn{} to the open-source community.

% 9. our contribution
We make the following contributions:
\begin{itemize}[noitemsep]
    \item Formulation of the schedule problem for distributed DNN training and inference given various operator placement strategies.
    \item Introduction of \pn{}, which is the first system to efficiently search for high-performance schedules based on the given operator placement strategies.
    \item Implementation of an end-to-end system that supports instantiating searched schedules for efficient runtime execution.
\end{itemize}

\section{Background and Motivation}
\label{sec:motivation}

\begin{figure}[t!]
\centering
    \includegraphics[width=\linewidth]{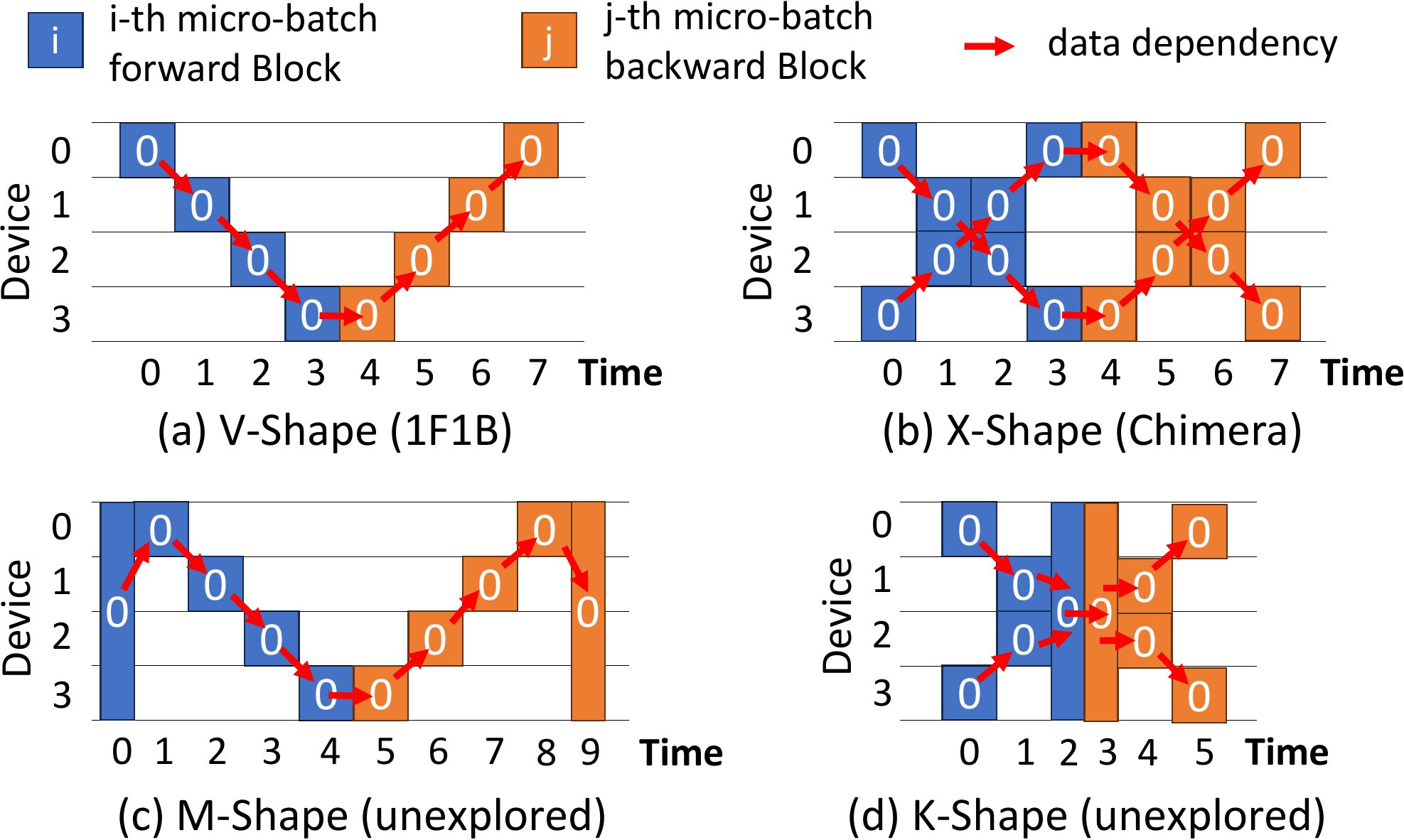}
    % \vskip -0.5ex
    \caption{Diverse possible operator placement strategies of one micro-batch. The blue and orange blocks denote forward and backward execution blocks, respectively. The red arrows denote data dependency between different blocks.
    % \TODO{replot (b) into 2 micro-batches instead of 1;}
    }
    % \vskip -2.5ex
    \label{fig:premise}
\end{figure}

% placment + schedule. well studied placement, pre-defined schedule
% \noindent 
For distributed DNN training and inference, previous works \cite{dapple, chimera, superscaler, branch-parallelism} have effectively explored flexible placement strategies for execution plans, showcasing their effectiveness in many real-world scenarios. Figure~\ref{fig:premise} uses a single micro-batch to illustrate several typical placement strategies, such as (a) grouping operators into execution blocks and sequentially placing them among devices~\cite{dapple}, (b) distributing them on distinct devices to form a bi-directional pipeline execution~\cite{chimera}, (c) employing a more advanced strategy to distribute memory-intensive operators across all devices~\cite{superscaler}, or (d) leveraging model architecture, \eg 2-branch, to place operators from independent branches on different devices~\cite{branch-parallelism}.
However, all these solutions employ pre-defined schemes such as 1F1B~\cite{dapple, pipedream} and Chimera~\cite{chimera} to schedule multiple micro-batches,
% for high device utilization, 
leaving the temporal schedule largely unexplored.
% \TODO{any supportive evidence, e.g., number of possible schedules?}
%
% \subsection{Limitations and Opportunities}
% \label{sec:motivation:limit}

\begin{figure}[t!]
\centering
    \includegraphics[width=0.85\linewidth]{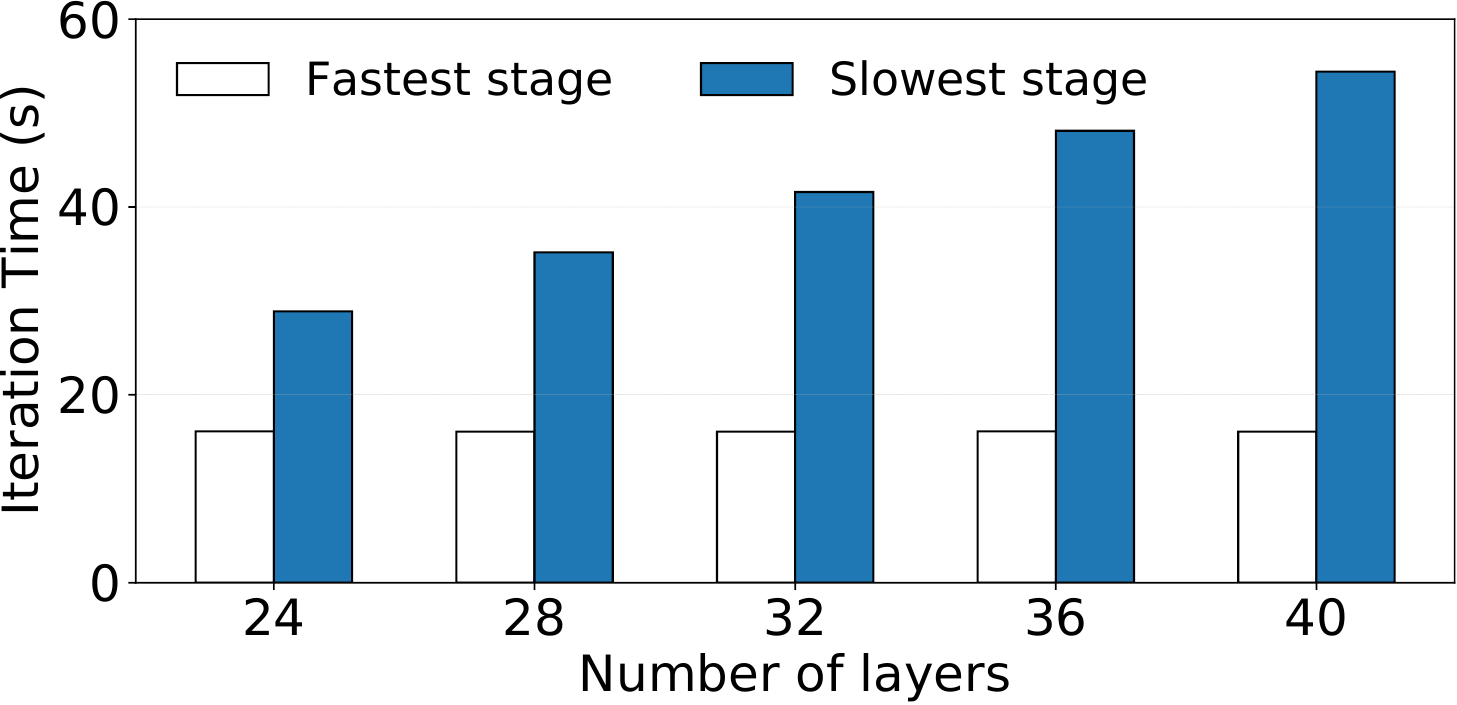}
    \vskip -0.5ex
    \caption{Training performance of GPT model with different number of layers. (Embedding vocabulary size 768k, layer from GPT~6.7B, on 4 V100-32GB GPUs).}
    % \caption{Micro-benchmark of training a 6.7B GPT model with an embedding layer containing 768k vocabulary size on 4 V100-32GB GPUs.}
    % \vskip -3.5ex
    \label{fig:microbench-moti}
\end{figure}

% diversity in dnn models pose new challenges
Furthermore, the emergence of large DNN models has introduced a significant increase in diversity, such as large embedding layers~\cite{gpt-4, vocab500k} in multilingual models, and multiple independent branches in multi-modal models~\cite{flava, simvlm, vilt}. Such diversity poses challenges to existing predefined schedules but, also presents new opportunities to further improve performance.
%
% using contentional placement + 1f1b schedule has low performance
Figure~\ref{fig:microbench-moti} shows the training performance of a GPT model with a large embedding layer using the 1F1B schedule following the Piper~\cite{piper} policy. With the increasing number of layers, computations on each device become increasingly imbalanced. The slowest stage is 3.4$\times$ slower than the fastest stage for the 40-layer GPT, leading to significantly lower device utilization in the first stage. This is because the large embedding layer consumes a significant amount of memory but requires only little computation cost. The large embedding layer requires at least two GPUs to fit in, leaving little room for co-locating other computation-intensive layers, \ie transformer~\cite{attention} layers. As a result, many computation-intensive layers can only be placed on the remaining two GPUs, leading to computational imbalance and performance drop.

% using advanced placement + 1f1b schedule need adjust, still low performance
A more advanced operator placement strategy such as the one shown in Figure~\ref{fig:premise}(c) may better fit this model, where the large embedding layer is distributed across all devices, allowing computation-intensive layers to share devices with the embedding layer. This can help alleviate the memory bottleneck, achieving a more balanced stage computation cost. However, predefined schedules such as 1F1B face challenges in direct application as it conflicts with the assumption that different stages are placed on distinct devices. While it's possible to manually adapt the 1F1B schedule to accommodate this placement strategy (\S\ref{sec:eval}),
% ill-considered 
these schedules would still suffer from low device utilization due to factors such as data dependency waiting. % \TODO{do we need a figure to demonstrate the ``inefficiency"?}
% A possible solution is to adopt a more advanced operator placement strategy, shown in Figure~\ref{fig:premise}(c), where the large embedding layer is distributed across all devices, allowing computation-intensive layers to share devices with the embedding layer. In this way, the memory bottleneck can be alleviated and the stage computation cost can be more balanced. However, this fails to directly apply 1F1B schedule as it assumes the stages are placed on distinct devices. Although one could manually adapt 1F1B schedule to accommodate with this placement strategy (\S\ref{sec:eval}), it still suffers from low device utilization due to data dependency waiting. % \TODO{do we need a figure to demonstrate the ``inefficiency"?}

% diversity + new placement call for efficient scheuldes.
The diversity in model-aware operator placements also highlights the importance of tailored schedules. Multi-modal models~\cite{flava, vilt}, for instance, treat distinct input modalities as separate branches, enabling a new placement strategy (\ie Figure~\ref{fig:premise}(d)) that concurrently executes each branch on different devices to reduce latency. However, there is no out-of-the-box schedule for such placement strategy,
making it necessary to find corresponding schedules.
% . Consequently, it's necessary to find efficient schedules specifically for these diverse operator placement strategies.
% to make them available.

% diversity brings more opportunities
% In addition to this, recent large models also consider handling multiple types of input modalities~\cite{flava}, such as images, texts and audio, by treating each modality as an independent branch that is combined into a single model. The output of each branch is then jointly considered to create a unified representation. As shown in Figure~\ref{fig:premise}(d), these models present opportunities to further speed up the inference of a single micro-batch by concurrently executing each branch on different device. However, existing schedule algorithms are designed based on sequential placement, and thus cannot adapt to new ones that leverage this type of operator placement. This limitation emphasizes the need for automatic search of efficient scheduling plans for diverse micro-batch execution plans. 

% \section{Problem Formulation}
\section{Problem Formulation And Insights}

\begin{table}[t!]
\normalsize
\begin{center}
\begin{tabular}{c|p{6cm}}
    Notation &  Meaning \\
    \hline \hline
    $N$     & Number of micro-batches \\
    $D$     & Number of devices \\
    $M$     & Memory capacity of each device \\
    $K$     & Number of blocks in each micro-batch \\
    $B_i^n$ & $i$-th execution block on $n$-th micro-batch \\
    $\mathbb{B}$ & Set of all blocks in a schedule \\
    $t_B$   & Time cost of executing block $B$ \\
    $s_B$   & Start time of executing block $B$ \\
    $d_B$   & Device(s) to execute block $B$ \\
    $m_B$   & Memory cost of executing block $B$ \\
    $B_i \rightarrow B_j$  & Data dependency where block $B_j$ depends on block $B_i$ \\
    \hline
\end{tabular}
\end{center}
\vskip -1ex
\caption{Notations used in this paper.}
\vskip -1ex
\label{tab:notation}
\end{table}

\noindent To better understand the schedule problem, we first formulate it as the problem of determining an optimal schedule for distributed DNN training or inference.
% micro-batches. 
We then highlight the challenges posed by the large search space inherent in this problem, which makes naive exhaustive search infeasible. 
% Finally, we present our key insights aimed at reducing the search space while still enabling the discovery of efficient schedules.
Finally, we present our key insights into reducing the search space during efficient schedules discovery.

\subsection{Problem Definition}
\label{sec:prob-def}

Consider an iteration of DNN model training that comprises $N$ independent micro-batches. The DNN model is distributed across $D$ homogeneous devices, each with the same memory capacity of $M$. The operator placement for each micro-batch is predetermined, and the execution of each micro-batch $n (0 \le n < N)$ can be represented as the execution of $K$ blocks denoted as $B_i^n$, where $0 \leq i < K$. Each block corresponds to a sub-set of operators on a device or a group of devices (utilizing tensor parallelism). Figure~\ref{fig:premise} illustrates several possible scenarios of operator placement strategies.
Each block $B$ has an associated execution time $t_{B} \in \mathbb{Z}^+$ and memory consumption $m_{B} \in \mathbb{Z}$. 
Without losing generality, we use integers ($\mathbb{Z}$) to express both $t_{B}$ and $m_{B}$ to maintain compatibility with tools such as the Z3-solver~\cite{z3-solver}. Certain blocks, such as backward computation, may exhibit negative memory consumption, indicating the release of memory after execution. % \gb{do we support the memory optimization like re-computation and swap? how? or it's limitation?}
% \TODO{Without loss of generality and also better compatible with tools like Z3-solver~\cite{z3-solver}, we use integer ($\mathbb{Z}$) to express both $t_{B}$ and $m_{B}$. Certain blocks, \eg{} backward computation, may have negative memory consumption, indicating releasing memory after execution.}
% The negative value of memory consumption indicates releasing memory consumption (\eg backward computation). 
The execution of each block can be determined to start at time
% $s_{B} \in \{s \mid s \ge 0, s \in \mathbb{Z}\}$.
$s_B (s_B \ge 0, s_B \in \mathbb{Z})$.
Table~\ref{tab:notation} presents a summary of the notations used in this paper.

\para{Exclusive execution constraints.} As a common practice~\cite{dapple, chimera}, we adhere to the exclusive execution of blocks on each device. This means that each device executes only one block at a time, as the operators in large DNN models typically saturate the device.

% memory constraints
\para{Memory constraints.} An executable schedule must satisfy memory constraints.
%, ensuring that the execution plan can be accommodated within the available memory capacities.
%The order in which blocks are executed on each device becomes crucial for managing peak memory consumption.
During a typical training iteration, executing a micro-batch consists of both the forward and backward computation. The forward computation involves the creation of tensors, which in turn consumes memory. Conversely, the backward computation involves the release of previously created tensors, thereby freeing up memory space. As a result, the careful arrangement of block execution can have a significant impact on memory utilization, especially when dealing with a large number of micro-batches, which can amplify the consequences of sub-optimal arrangements.
% \TODO{especially when facing a large amount of micro-batches which will amplify the impact of mis-arragengment.}

% data dependency constraints
\para{Data dependency constraints.} 
Given the operator placement strategy, individual blocks typically exhibit data dependencies, meaning that the completion of prior dependent blocks is a prerequisite for the start of subsequent blocks. This ensures the correct flow of data and computations throughout the DNN models.
A valid schedule must strictly follow data dependency to preserve the semantics of the DNN model.
% \st{while simultaneously avoiding potential deadlocks.} \TODO{only following data-dependency cannot prevent deadlock}

\para{Objective.} The primary objective in exploring an efficient schedule is to minimize the total execution time by searching the start time of each block while adhering to the above constraints. The execution time can be quantified as the time required to complete the execution of the last block among all the blocks ($\mathbb{B}$) within the schedule. Thus, the optimization goal can be formulated as follows, taking into account the aforementioned constraints:

\begin{figure}[t]
    \centering
    \includegraphics[width=0.85\linewidth]{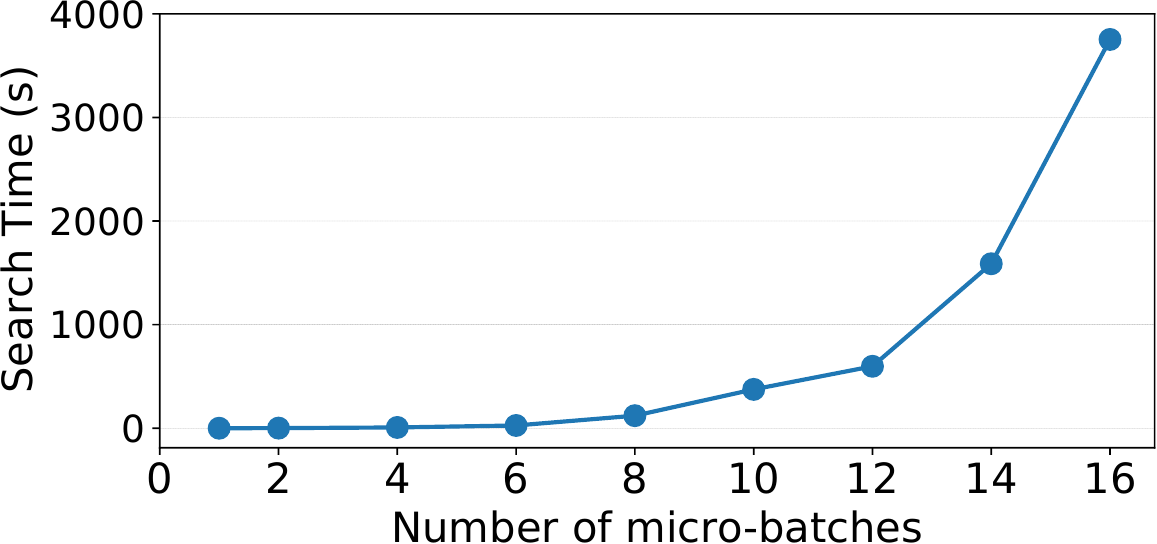}
    \caption{Schedule search time of V-Shape placement with different number of micro-batches. 
    % \TODO{a curve of $y=C^x$?}
    }
    \vskip -3ex
    \label{fig:moti-search-time}
\end{figure}

\begin{figure*}[t!]
\centering
    \includegraphics[width=\linewidth]{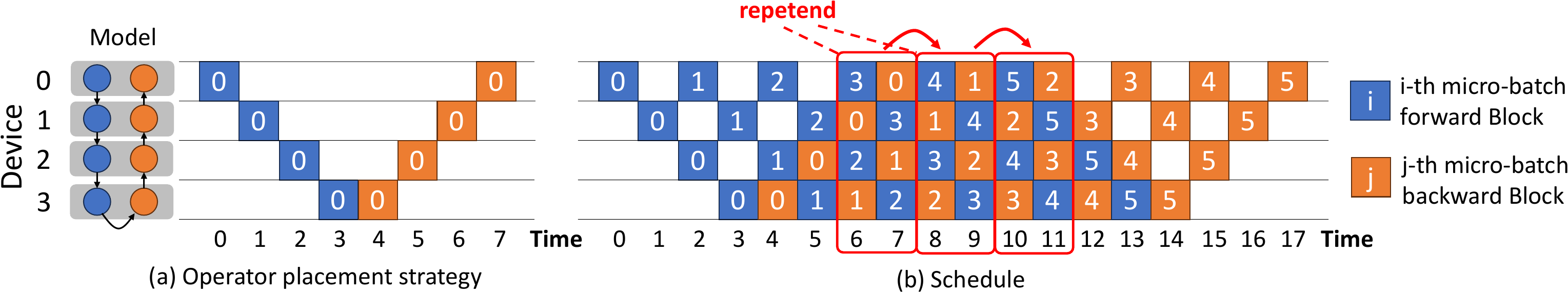}
    \vskip -0.5ex
    \caption{One micro-batch execution from (a) model operator placement, is extended into (b) 6 micro-batches with 1F1B schedule.
        % The blue and orange blocks stand for the forward and backward blocks, respectively.
    }
    \vskip -1ex
    \label{fig:repetend}
\end{figure*}

% \begin{equation}
% \begin{aligned}
% \underset{s}{\text{minimize}} \quad & \underset{B \in Blocks}{\max} s_B + t_B \\
% \text{subject to} \quad & \text{(1)} \quad s_{B_i} + t_{B_i} \le s_{B_j}, \quad \forall B_i, B_j \text{ if } B_i \rightarrow B_j \\
% & \text{(2)} \quad \underset{d \in K}{\max} \text{ PeakMem}(\{ m_{B_i} \mid d \in d_{B_i} \}) \le M \\
% \end{aligned}
% \label{eq:prob-def}
% \end{equation}

%\begin{equation}
%\begin{aligned}
% \underset{s}{\text{minimize}} \quad & \underset{B \in Blocks}{\max} s_B + t_B \\
% \text{subject to} \quad & \text{[1]} \quad s_{B_i} + t_{B_i} \le s_{B_j}, \quad \forall B_i, B_j \text{ if } B_i \rightarrow B_j \\
% & \text{[2]} \quad \underset{d \in D}{\max} \text{ PeakMem}(\{ m_{B_i} \mid d \in d_{B_i} \}) \le M \\

%\underset{\{s_{B}\}}{\text{argmin}} & \quad \underset{ B \in \{ B_i^n \mid i \in [ 1, K ], n \in [ 1, N ] \} }{\max} ( s_B + t_B ) \\
%\text{subject to}: \\
    %& \text{[1]} \; s_{B_i} + t_{B_i} \le s_{B_j}, \quad \forall B_i, B_j \text{ if } B_i \rightarrow B_j \\
    %                    & \quad \quad \quad \quad \quad \quad \quad \quad \quad  \text{ or } i \ne j \text{ and } d_{B_i} = d_{B_j} \\
    %& \text{[2]} \; M \ge \underset{ \substack{ d \in [1, D] \\ \tau \in [0, +\infty )}}{\max} ( \sum_{\{B_i^n \mid d_B = d, s_B \in [0, \tau), \forall i, n \}} m_{B_i^n} ) \\
%\end{aligned}
%\label{eq:prob-def}
%\end{equation}

\vspace{-2ex}
\begin{equation}
\begin{aligned}
\text{minimize} & \quad \underset{ B \in \mathbb{B} }{\max} \quad ( s_B + t_B ) \\
\text{subject to:} & \\
    \text{[1]} & \quad \min(s_{B_i} + t_{B_i}, s_{B_j} + t_{B_j}) \le \max({s_{B_i}, s_{B_j}}), \\
    & \quad \forall B_i, B_j \text{ if } i \neq j, d_{B_i} = d_{B_j} \\
    \text{[2]} & \quad M \ge \underset{ \substack{ d \in [0, D) \\ \tau \in [0, +\infty )}}{\max} ( \sum_{B \in \mathbb{B}_\tau^d}{m_{B}}), \\
    & \quad \text{where} \quad \mathbb{B}_\tau^d = \{B \mid d_B = d, s_B \in [0,\tau) \} \\
    \text{[3]} & \quad s_{B_i} + t_{B_i} \le s_{B_j}, \quad \forall B_i, B_j \text{ if } B_i \rightarrow B_j
\end{aligned}
\label{eq:prob-def}
\end{equation}

% The above \text{[2]} calculates the peak memory usage for a device during execution by firstly applying the cumulative sum~\cite{cumsum} operation on the memory of its blocks according to the execution starting time $s_{B_i}$, and then identifying the maximal value as the peak memory usage.

In the above, item~\text{[2]} calculates the peak memory of a device by cumulatively summing the memory of blocks following their execution order, with the maximum value considered as the peak memory. 
%
% The data dependency and memory constraints play crucial roles in determining the efficiency of various schedules. %~(\S\ref{sec:eval}).
% % \TODO{$\leftarrow$ not clear, do you mean ``presence $\rightarrow$ preserving''?}
% % These constraints have a direct impact on the overall performance of the schedules.
% For instance, a device may experience idle time while waiting for the arrival of required data due to data dependencies. Similarly, the execution of backward blocks might be prioritized over forward blocks to release memory due to memory constraints, leading to delays even when a forward block is ready for execution. Consequently, carefully considering
% % and effectively managing
% these constraints is essential to optimize the efficiency of schedules. 
% % Note the communication cost between concurrent blocks, \eg AllReduce of data-parallelism can be included in the block, 
% % while the communication between consecutive data-dependent blocks is  relatively small cost send-receive than block execution~\cite{alpa}. Therefore, we don't consider communication cost separately in schedule.
Note that the communication between data-dependent blocks incurs a relatively small cost compared to block execution~\cite{alpa}. Hence, we have omitted it in the schedule.
% \zhiqimodify{Note we don't consider the communication cost between blocks in the schedule since it is significantly smaller compared to the cost of executing the blocks themselves~\cite{megatron3}.}
% \TODO{looks communication cost omitted in this modeling, we need an explanation, e.g., 1) model partitioning already considered minimizing communication, which makes communication cost small when comparing with computation cost; 2) given model partitioning plan, communication pattern is fixed, so does the cost, then we included in the adjacent computation; 3) other systems also think this way, \eg Alpa?}

The schedule problem naturally supports the combination of existing tensor and data parallelisms~\cite{data-parallel, megatron1}, where operator placement strategies can determine the mapping of blocks to multiple devices for concurrent execution. Given the focus of this paper on efficient schedule searching, we have simplified the presentation of schedules in our figures. 

\subsection{Problem Space}

%\TODO{this paragraph will be moved to a later section as space reduction technique explanation $\rightarrow$}
%Given the computation equivalence among micro-batches, we observe that a schedule can always be alternatively represented as a performance-equivalent schedule, where the micro-batch indices monotonically increase for every same block $B_i$ with growing time steps. Therefore, we search schedules that adhere to this characteristics throughout the paper.

% Clearly, \TODO{$\leftarrow$ need to explain} 
This type of schedule problem is known to be NP-hard~\cite{nphard}. The complexity of the problem grows exponentially as the number of micro-batches increases,
% \zhiqimodify{
due to the independence of blocks among different micro-batches, \ie

\vspace{-2ex}
\begin{equation}
\begin{aligned}
B_i^m \nrightarrow B_j^n, \quad \forall i, j \in [0, N) \text{ if } m \neq n
\end{aligned}
\end{equation}

\noindent Such independence leads to the possibility of arbitrary execution order on these blocks for each device.
% }
In practical scenarios, where the number of micro-batches can range from hundreds to thousands~\cite{megatron2}, the search space becomes prohibitively large to explore using brute-force methods.

To illustrate this challenge, we begin with a GPT model and follow the conventional approach of sequentially placing its operators onto 4 devices, as depicted in Figure~\ref{fig:premise}(a). We employ the Z3-solver~\cite{z3-solver} to encode and solve the schedule problem, aiming to identify the optimal schedule as the number of micro-batches increases. For simplicity, we assign an execution time of 1 to each forward block and 2 to each backward block.
Figure~\ref{fig:moti-search-time} shows the results of the search time with an increasing number of micro-batches. It is evident that as the number of micro-batches grows, the search time increases significantly. In fact, it takes 3752 seconds to complete the search for only 16 micro-batches, indicating that it becomes impractical to search for the optimal schedule when dealing with hundreds or even thousands of micro-batches.

Consequently, it is necessary to address the challenges brought about by the large search space.

% \if 0
\subsection{Key Insights}
\label{sec:prob-insight}

% \TODO{the insight subsection already beyond the definition of ``problem formulation''}

% 1) we can identify a \textbf{repetend} in a small schedule plan that can generalize the plan to any number of micro-batches 2) furthermore, the repetend can be constructed with property

By carefully checking many efficient schedules, we observed that:
% made observations that provide insights in solving the problem.
1) efficient schedules usually exhibit repetitive cycles, wherein the same computations are periodically performed 
but over different micro-batches;
% as the micro-batch indices increase.
2) repetitive cycles, referred to as \textbf{repetend}, 
only involves a small number of micro-batches;
% can be identified and constructed within a small number of micro-batches. 
3) the repetend occupies most of the time, especially when the number of micro-batches is large, with a small warmup phase in the beginning and a cooldown phase in the end.

\para{Repetend.}
Figure~\ref{fig:repetend} demonstrates the 1F1B schedule on 4 devices. The red boxes show 3 repetends, with the same execution repetitively performed.
For example, comparing steps 8-9 with steps 6-7, each device executes the same blocks, except with every micro-batch index increased by one.
% Such execution pattern repeats until step 12, where no more micro-batches need to start computing. % Such plan can be generalized to any number of micro-batches for real practice.

% Furthermore, the schedules identified through this approach can be generalized to any large number of micro-batches.

% \para{Repetend.} Figure~\ref{fig:repetend} demonstrates the well-designed 1F1B schedule on 4 devices. As shown in the red boxes, we found the schedule repeatly performs the same execution with increasing micro-batch number. For example, comparing step 8-9 to the step 6-7, every device executes the same block with increasing one micro-batch index. Such execution pattern repeats until step 12, where no more micro-batches need to start computing. % Such plan can be generalized to any number of micro-batches for real practice.

\para{Schedule generalization.} 
% The presence of a repetend introduces a property wherein the schedule can be extended to accommodate any number of micro-batches larger than the schedule itself. 
We found that if the micro-batch indices between consecutive repetends increase by exactly one,
it is possible to extend the repetend schedule to accommodate any number of micro-batches.
% \TODO{if a repetend with micro-batch index increased by 2, then cannot be extended for ``any number of micro-batches'', e.g., 7-micro-batches}
To illustrate this, consider extending the 6-micro-batch 1F1B schedule in Figure~\ref{fig:repetend} to 7 micro-batches. This can be achieved by replicating the repetend of steps 10-11 while increasing the micro-batch indices of each block in the repetend by one. % Consequently, 
The blocks originally following step 12 would be shifted by the width of the repetend accordingly, with each block also increasing its micro-batch index by one.
%
% By employing this approach, we can formulate a general schedule consisting of three components: warmup, repetend and cooldown. The warmup refers to the blocks that happen before the start of the repetend, while the cooldown consists of the blocks that start after the end of the repetend. %To scale the plan to accommodate any desired number of micro-batches, the repetend can be repetitively applied with each repetition increasing the micro-batch index by one, until the desired number of micro-batches is reached. Correspondingly, the cooldown blocks increase their micro-batch index in accordance with the number of repetitions.
% Therefore, to scale the schedule to accommodate any desired number of micro-batches, such extension processing can be applied repetitively.
% the repetend can be repetitively applied with increasing micro-batch indices.

Based on these observations, we 
can simplify the problem by transforming the search for a full schedule for all micro-batches into a repetend search that involves much fewer micro-batches. 
% can effectively identify efficient schedules by addressing a reduced-scale problem.
Subsequently, it can schedule any large number of micro-batches by extending the efficient repetend and incorporating a proper warmup and cooldown phase schedule.

% \para{Conditions to be a repetend.} Through further investigation, inside a schedule, we have identified the sufficient condition for a repetend to exist: a repetend must consist of blocks that collectively perform a complete computation of one micro-batch, irrespective of their micro-batch indices. Formally, a repetend exists in a scheduling plan when there is a range of steps from $i$ to $j$ in which the blocks contain:

% \begin{equation}
% \begin{aligned}
% \{B_k \mid i \le s_{B_k} \le j \} = \{B_1, B_2, ..., B_K\}
% \TODO{unique}
% \end{aligned}
% \end{equation}

% For example in Figure~\ref{fig:repetend}, step 6-7 is a repetend which contains a complete computation of the one micro-batch, disregarding the micro-batch index associated with each block. 
% \TODO{Due to space limitation, we skip the proof of this property.}
% This observation allows us to conduct a quick search by first identifying a small schedule with a repetend and then generalizing it to accommodate the desired number of micro-batches.

% \fi

% \TODO{To substantiate this observation, we provide a comprehensive proof of this property in the Appendix}.
\section{Design}
\label{sec:design}

\noindent Following the above insights, we present \pn{}, a system that supports automated search for efficient schedules given operator placement strategies.

\subsection{\pn{} Overview}

\begin{figure}[t!]
\centering
    \includegraphics[width=\linewidth]{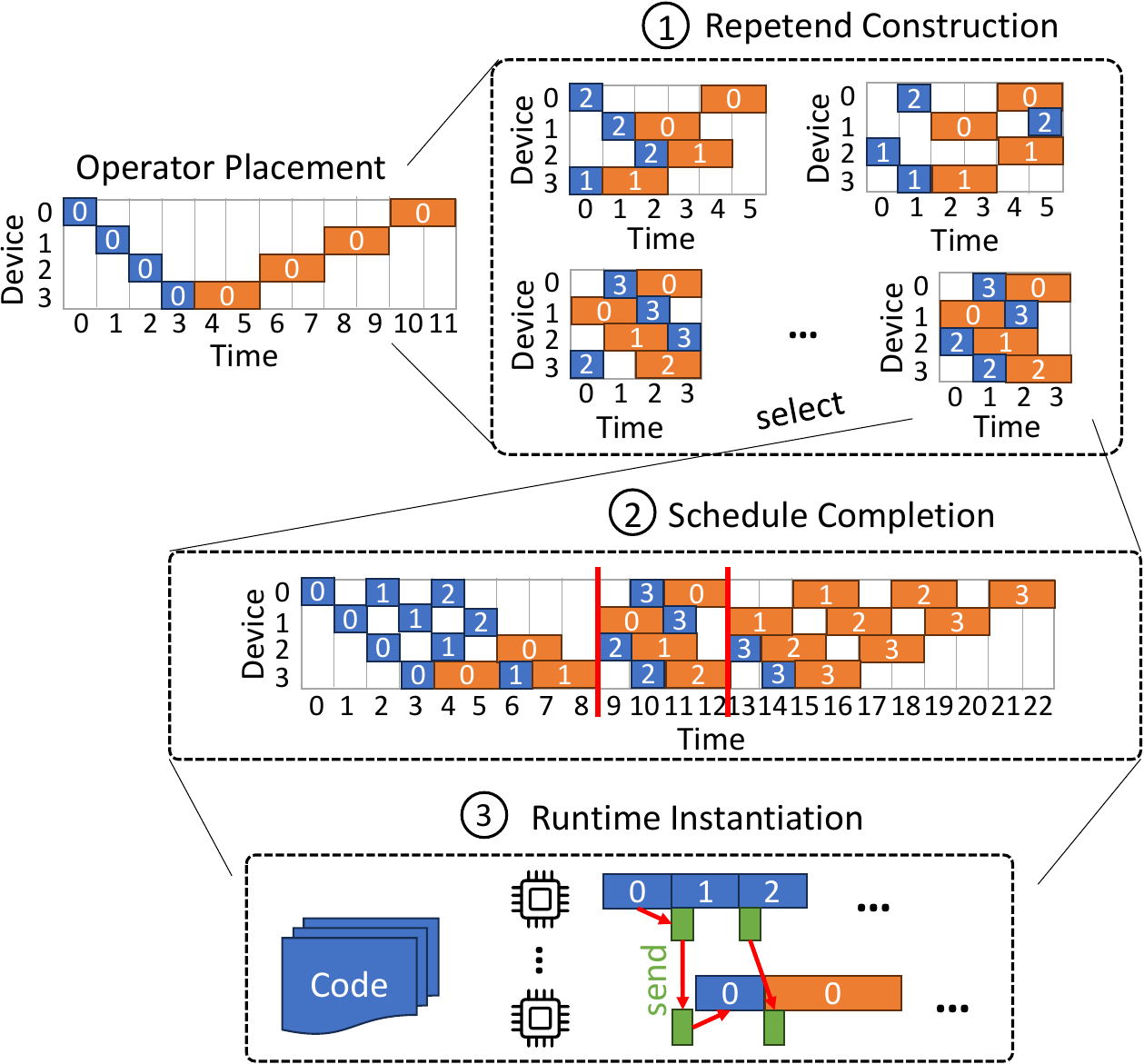}
    % \vskip -0.5ex
    \caption{\pn{} overview.}
    % \vskip -1ex
    \label{fig:overview}
\end{figure}

Figure~\ref{fig:overview} illustrates the overview of \pn{}. \pn{} takes the operator placement strategy and memory budget as inputs. Within the system, \pn{} employs three phases for an end-to-end automated search and execution: repetend construction, schedule completion and runtime instantiation, where the first two phases search for an efficient schedule and the last phase instantiates the schedule to runtime for real execution.

% \zhiqi{may move to another place}

% \zhiqimodify{With the same operator placement strategy among micro-batches, we observe that a schedule can always be alternatively represented as a performance-equivalent one, where the micro-batch indices monotonically increase for every same block $B_i$ with growing time. Therefore, we search schedules that adhere to this characteristics throughout the paper.} \TODO{independent - symmetry - ordering} \zhiqi{check whether we can make this more intuitive.}

Given an operator placement strategy, in repetend construction, \pn{} firstly samples all possible blocks from a small number of micro-batches ($N_R$) that can form a repetend, and then picks the one with the lowest execution time. Then, given the selected repetend, \pn{} completes its warmup and cooldown phase by searching for the time-optimal schedule on the remaining blocks, and extends the schedule to the desired number of micro-batches ($N$). Finally, according to the searched schedule, \pn{} inserts communication primitives and generates executable code for each device for runtime. % execution.

\subsection{Repetend Construction}
\label{sec:design-construct}

% leading paragraph:
% 1) repetend dominates the performance
% 2) the problem space
Given the substantial number of micro-batches, 
%which can range from hundreds to thousands,
the repetend is repeated numerous times 
and covers the major portion of iteration time, 
thus dominating performance.

\para{Block set of repetend.} Through further investigation, % inside a schedule,
we identified a necessary condition for a repetend: a repetend must consist of 
a full set of blocks in the model,
% blocks that collectively perform a complete computation of one micro-batch, 
irrespective of their micro-batch indices.
For example in Figure~\ref{fig:repetend}, step 6-7 is a repetend that contains 
% a
all blocks to complete the
computation of a micro-batch, disregarding the 
micro-batch indices associated with the blocks.
Therefore, blocks of a repetend can be expressed as:
%\footnotemark:
% Therefore, a repetend existing in a schedule consists of blocks $\mathbb{B}_{repetend}$:

\begin{equation}
\begin{aligned}
    \mathbb{B}_{repetend} &= \{B_0^{n_0}, B_1^{n_1}, ..., B_{K-1}^{n_{K-1}}\}, \\
    % \text{where} \quad {m_1, ..., m_S} &\in \{1, ..., S\}
    % \text{where} \quad \{{m_1, ..., m_S}\} &= \{1, ..., S\}
    % \text{where} \quad {n_1, ..., n_K} &\in \{1, ..., N\}
    \text{where} & \quad {n_0, ..., n_{K-1}} \in [0,N_R)
\label{eq:repetend}
\end{aligned}
\end{equation}

% \TODO{Due to space limitation, we skip the proof of this property.}
% This observation allows us to conduct a quick search by first constructing a repetend from a small number of micro-batches ($N_R$) and then generalizing it to accommodate the desired number of micro-batches.

% picking repetends:
% 1) construction space is large, each B_i can be assigned to any micro-batch indices
% 2) property-2: dependency leads to micro-batch index pruning
% 2) insights: the number of largest micro-batch index decides the peak memory usage
% 3) earch repetends composing from smaller micro-batch number to large micro-batch number

% \para{Repetend selection.} When dealing with a large number $N$ of micro-batches, each execution block $B_i$ has the flexibility to be assigned by any micro-batch index ranging from $[0,N_R)$. Consequently, there exist a maximum of $(N_R)^K$ potential repetends, resulting in a large space.

\para{Space pruning.} Constructing a repetend from $N_R$ micro-batches leads up to $(N_R)^K$ potential repetends,
which still presents a large space. 
Fortunately, we identified two key properties
% \footnotemark[1]\footnotetext[1]{Proofs are omitted in this draft due to space constraints.}
that can significantly reduce the search space:

\vspace{0.5ex}
\begin{property}
% $\forall B_i \in \mathbb{B}, m < n \text{ if } s_{B_i^m} < s_{B_i^n}$
% $\forall B_i^m, B_i^n \in \mathbb{B}, m < n \text{ if } s_{B_i^m} < s_{B_i^n}$.
% $\forall \mathbb{B}, \exists \mathbb{B'}, s.t. \text{perf}(\mathbb{B}) = \text{perf}(\mathbb{B'}), \forall B_i^m, B_i^n \in \mathbb{B'}, s_{B_i^m} < s_{B_i^n} \Rightarrow  m < n$.
For any schedule, there exists a same-performance schedule of blocks $\mathbb{B}$, $\forall B_i^m, B_i^n \in \mathbb{B}, m < n \text{ if } s_{B_i^m} < s_{B_i^n}.$
\label{prop:symmetric}
\end{property}

\vspace{1ex}
\begin{property}
$\forall B_i^m, B_j^n \in \mathbb{B}_{repetend}$, $m \ge n$ if $B_i^0 \rightarrow{} B_j^0$.
\label{prop:repetend}
\end{property}

\vspace{1ex}
Property~\ref{prop:symmetric} exploits the symmetry of all micro-batches, such that switching micro-batch indices for blocks doesn't affect the performance of the schedule. Thus, we can deduplicate symmetric schedules by focusing only on those where micro-batch indices monotonically increase over time for each block.

Property~\ref{prop:symmetric} further leads to Property~\ref{prop:repetend}. Within a repetend, we discovered that if $B_i$ has data dependency on $B_j$ inside one micro-batch, the micro-batch index assigned to $B_i$ should not be smaller than the one assigned to block $B_j$. This enables a pruning strategy, constraining the assignment of micro-batch indices to a sequence of dependent blocks in descending order.

\begin{figure}[t!]
\centering
    \includegraphics[width=0.95\linewidth]{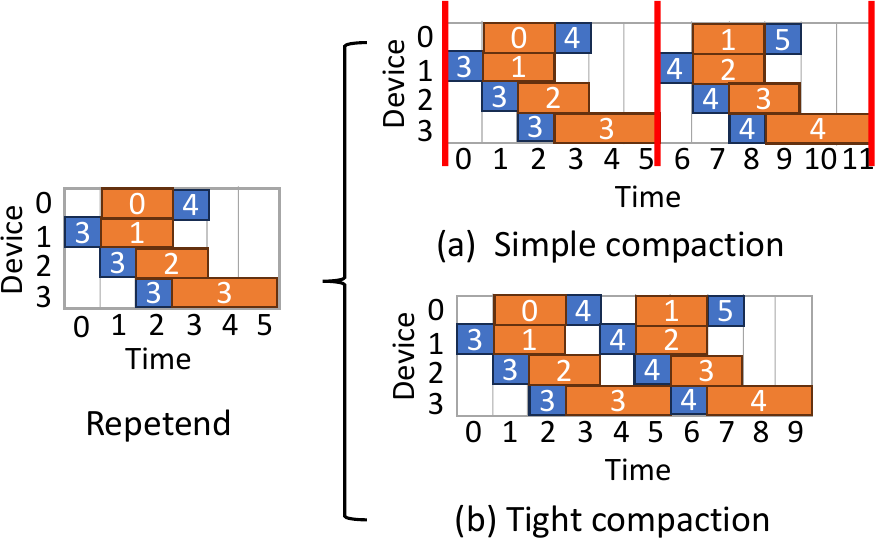}
    \caption{Two ways of compacting neighboring repetends. The repetend is based on a V-Shape operator placement.}
    % \vskip -1ex
    \label{fig:repetend-compact}
\end{figure}

In addition to the pruned space, Property~\ref{prop:symmetric} indicates that larger $N_R$ can potentially lead to larger peak memory usage.
As illustrated in the left of Figure~\ref{fig:repetend-compact}, device~0 needs to execute 4 forward blocks ($B^0 \sim B^3$) prior to this repetend, while no backward block is executed since the repetend includes the first backward block with micro-batch~0.
% \gb{you mean the backward block with micro-batch index = 0 in device0?}
Given the limitation imposed by peak memory usage, our search strategy begins with a small number of micro-batches ($N_R)$, and gradually increases it until 
% the lower bound of the execution time reached or 
we reach the memory capacity constraints.

% performance and evaluate metric:
% 1) different performance of repetend is due to different idle time, caused by data dependency and memory constraints;
% 2) repeating repetend have opportunity to further reduce the idle waiting time
% 3) based on 1) and 2) we use use execution time to evaluate a repetend performance, and introduce the algorithm to consider a more efficient concatenation
\para{Repetend performance.} 
Different repetends lead to varying execution performance, as illustrated in Figure~\ref{fig:overview}~\textcircled{\footnotesize 1}.
% The repetend construction phase in Figure~\ref{fig:overview} already demonstrated different repetends that lead to varying execution performance. 
Their performance mainly differs in device idle time, which can be attributed to the presence of data dependency and memory constraints.
A block can only start after all its data-dependent blocks are finished,
% And the memory constraints necessitate the executing the backward block before the forward block, when there is insufficient memory to execute another forward block until a backward block is finished to release memory. 
and the memory constraints might require a device to wait until a backward block releases memory, even if other forward blocks are ready for execution.

% These constraints can lead to idle time for the device, resulting in sub-optimal schedule efficiency.
% The observed variations in performance can be attributed to the presence of data dependency and memory constraints, \eg the memory constraints necessitate executing the backward block before the forward block, as there is insufficient memory available for executing a forward block until a backward block is executed to release memory. These constraints can lead to idle time for the device, resulting in sub-optimal schedule efficiency.

% Specifically, in Figure~\ref{fig:repetend-perf}(a), the start of the block on device 1 is delayed until the completion of the block on device 0 due to data dependency. Additionally, in Figure~\ref{fig:repetend-perf}(b), the memory constraints necessitate executing the backward block before the forward block, as there is insufficient memory available for executing a forward block until a backward block is executed to release memory. These constraints can lead to idle time for the device, resulting in sub-optimal scheduling efficiency.

% The performance of a repetend can be evaluated by measuring
% the execution time of a repetend,
% which is dominated by the device with the longest execution time.
% , where lower execution time indicates higher performance. 
% The performance of a repetend is constrained by the device with the highest execution time, which determines the overall efficiency. 
% Initially, the execution time can be estimated by the time span from the start execution time of the first block to the end of finish execution time of the last block. 
Considering the idle time, the execution time of a repetend can be easily evaluated from its first block to the last one.
However, this estimation is not precise when certain idle time slots at the end of a repetend can be utilized by a frontier block in the subsequent repetend. 
Figure~\ref{fig:repetend-compact} demonstrates this case. In Figure~\ref{fig:repetend-compact}(a), the repetend is repeated immediately after the completion of the previous repetend. However, we can optimize this pattern by initiating the next repetend as soon as the dependent blocks are finished, rather than waiting for the entire previous repetend to finish. Figure~\ref{fig:repetend-compact}(b) shows this approach, where the subsequent repetend starts at time slot 4,
%\gb{slot 5?}
as it no longer needs to wait for the completion of the previous repetend due to the timely completion of its dependent blocks by slot 4. This results in a tighter compaction between repetends, effectively reducing device idle time and enhancing overall schedule efficiency.

% To provide a more accurate evaluation of repetend performance, we enhance the previous estimation by incorporating the above overlapping nature of repeated repetends. 
Therefore, we incorporated the above-mentioned overlapping nature to evaluate repetend performance more accurately,
% The execution time of a repetend can be decomposed
decomposing its time into execution time $E_R$ and wait time $W_R$. On a device, execution time refers to the duration from the start of the execution of the first block to the completion of the last block, with potential device idle time incorporated. On the other hand, wait time corresponds to the idle time between two consecutive repetends.
% , spanning from the completion of the last block's execution in the previous repetend to the initiation of the first block's execution in the subsequent repetend on that device. 
The efficiency of a repetend determined by the device with the longest time. Therefore, the overall repetend performance ($t_R$) can be summarized as:

\vspace{-1ex}
\begin{equation}
\begin{aligned}
t_R &= \max_{d \in [0,D)} (E_R^d + W_R^d), \\
\text{where} \quad E_R^d &= \max_{B_i \in \mathbb{B}_d} (s_{B_i} + t_{B_i}) - \min_{B_i \in \mathbb{B}_d} s_{B_i}, \\
\text{where} \quad \mathbb{B}_d &= \{ B_i \mid d_{B_i} = d \} \\
\end{aligned}
\end{equation}

$W_R$ can be calculated by traversing the execution blocks of the succeeding repetend and determining their earliest starting time based on their between-block data dependencies.
% Details of the calculation have been omitted in this paper due to space limitations.
% While the specific algorithmic details are not presented in this paper due to space limitations, the calculation of $W_R$ involves assessing the timing constraints imposed by the dependencies between blocks within the repetend.

% memory constraints
% minimize bubble rate using z3 solver:
% 1) we use Z3 solver to encode the problem
% 2) memory information can know before the repetends
% \para{Minimizing repetend execution time.} 
\para{Efficient repetend search.} 
% Given the assignment of micro-batch indices to each execution block, there are various of scheduling choices to construct a repetend. 
To find the optimal schedule of the repetend, \ie the shortest execution time of repetend $t_R$, we leveraged Z3-solver~\cite{z3-solver} to encode the schedule problem and minimize $t_R$~(\S\ref{sec:imple}). 
For the memory constraints, 
% by analyzing the blocks assigned with micro-batch indices, 
by analyzing the micro-batch indices assigned to the blocks in repetend,
we can learn the blocks that need to be executed prior to the repetend, and infer the memory usage at the entry of the repetend.
% As a result, the memory usage at the entry of the repetend can be inferred. 
Based on this information, we can set the memory constraints for the repetend during the search.
% Consequently, during the search process, the memory constraints can be updated based on this information. % \zhiqi{check whether the ``memory usage'' here is mis-leading}

\subsection{Schedule Completion}
\label{sec:design-completion}

% \para{Schedule generalization.} The presence of a repetend introduces a property wherein the schedule can be extended to accommodate any number of micro-batches larger than the plan itself. To illustrate, consider extending the 1F1B schedule depicted in Figure~\ref{fig:repetend} to containing 6 micro-batches. This can be achieved by replicating the repetend of step 10-11 while increasing the micro-batch index by one for each block inside the repetend. Consequently, the blocks originally after step 12 would be shifted accordingly, with each block also increasing its micro-batch index by one.

% By employing this approach, we can formulate a general schedule consisting of three components: warmup, repetend and cooldown. The warmup refers to the blocks that happen before the start of the repetend, while the cooldown consists of the blocks that start after the end of the repetend. %To scale the plan to accommodate any desired number of micro-batches, the repetend can be repetitively applied with each repetition increasing the micro-batch index by one, until the desired number of micro-batches is reached. Correspondingly, the cooldown blocks increase their micro-batch index in accordance with the number of repetitions.
% To scale the schedule to accommodate any desired number of micro-batches, the repetend can be repetitively applied with increasing micro-batch indices.

% leading paragraph:
% 1) need to complete the warmup and cooldown
% 2) what blocks needed for composition of warmup and cooldown
To create a complete schedule, we need to include the warmup and cooldown phases alongside the repetend. 
% Both phases are necessary to ensure the proper initialization and conclusion of the schedule.
Specifically, if the repetend involves a total of $N_R$ micro-batches, for any block $B_i^r$ within the repetend ($\mathbb{B}_{repetend})$, the warmup phase should comprise the blocks defined as:

\vspace{-2ex}
\begin{equation}
% \text{Blocks} = \{ B_i^m \mid 1 \leq m < r \}
\mathbb{B}_{warmup} = \{ B_i^n \mid \forall B_i^r \in \mathbb{B}_{repetend}, 0 \leq n < r \}
\end{equation}

% This contains all blocks of $B_i^m$ with micro-batch indices ranging from 1 to $r$ (exclusive). 
Similarly, the cooldown phase should include the blocks:

\vspace{-2ex}
\begin{equation}
% \text{Blocks} = \{ B_i^m \mid r < m \leq N_R \}
\mathbb{B}_{cooldown} = \{ B_i^n \mid \forall B_i^r \in \mathbb{B}_{repetend}, r < n < N_R \}
\end{equation}

% Here, the blocks $B_i^m$ with micro-batch indices greater than $r$ up to $N_R$ (the total involved number of micro-batches in repetend) are part of the cooldown phase.
% By incorporating necessary warmup and cooldown parts, the resulting schedule becomes comprehensive to scale with any number of micro-batches larger than $N_R$.

% step optimal search:
% 1) equation for step optimal 
% 2) memory constraints
\para{Time-optimal search.} We also employ Z3-solver to perform an optimal search for both the warmup and cooldown phases independently. The objective of this search is to minimize the execution time. The same objective and constraints of this search refer to Equation~\ref{eq:prob-def}. Note the memory constraints in the cooldown phase will be adjusted accordingly given the execution blocks in the warmup and repetend phases.
% By treating the warmup and cooldown phases as separate optimization problems, we can efficiently explore and identify the most efficient solutions for each phase.

\begin{algorithm}[t!]
\DontPrintSemicolon
   \small
   \KwInput{Operator placement strategy $OPS$}
   \KwInput{Memory constraints $M$}
   \KwOutput{General Schedule $schedule$}

   \tcp{init the upper bound of repetend time}
   optimal $\gets 0$ \\
   \For{\upshape $B_i \in$ $OPS$.blocks}
   {
      optimal $\gets$ optimal + $t_{B_i}$ \\
   }

   schedule $\gets$ None \\
   lower\_bound $\gets$ GetLowerBound($OPS$) \\

   inflights $\gets$ CalMaxInflight($OPS$, $M$) \\

  \For{\upshape $N_R \gets$ 1 to inflights}
  {    
    \For{\upshape repetend $\gets$ IterRepetendBlocks($OPS$, $N_R$)}
    {
        $sched_R \gets$ RepetendSolver(repetend, $M$) \\
        $t_R \gets$ RepetendTime($sched_R$) \\

        \uIf{\upshape $t_R \ge optimal$}
        {
          \textbf{continue}
        }
        optimal $\gets$ $t_R$ \\
        \tcp{get warmup and cooldown blocks}
        warmup $\gets$ GetWarmupBlocks(repetend, $N_R$) \\
        cooldown $\gets$ GetCooldownBlocks(repetend, $N_R$) \\

        \tcp{solver for time-optimal plan}
        $sched_W \gets$ TimeOptimalSolver(warmup, $M$) \\
        $sched_C \gets$ TimeOptimalSolver(cooldown, $M$) \\

        \tcp{compose to the general plan}
        $schedule$ $\gets$ Concat($sched_W$, $sched_R$, $sched_C$) \\

        \tcp{early exit}
        \uIf{\upshape $t_R$ = lower\_bound}
        {
            \Return $schedule$
        }
    }
  }
  \Return $schedule$

\caption{\pn{} schedule search.}
\label{alg:search-overview}
\end{algorithm}

% pesude-algorithm
\para{Putting it altogether.} Algorithm~\ref{alg:search-overview} shows the 
overall % complete process of the 
search algorithm, given the operator placement strategy $OPS$ and memory constraints $M$ as inputs. 
% \st{As the repetend plays a crucial role in determining the efficiency of the overall plan, the objective of the search algorithm is to discover a schedule with the shortest repetend execution time.}
The algorithm begins by initializing the upper bound (\texttt{optimal}) and lower bound of the repetend execution time (Lines 1-5). The function \texttt{GetLowerBound} computes the lower bound of the repetend by summing the execution times of the blocks assigned to each device of one micro-batch and selecting the maximum value across all devices. Additionally, based on the operator placement strategy, the algorithm determines the maximum number of in-flight micro-batches that can be executed within the memory constraints (Line 6).

The search algorithm then follows the strategy outlined in \S\ref{sec:design-construct}. It initiates the search by considering a smaller number of micro-batches $N_R$ within the repetend (Line 7). The algorithm then iterates over all possible assignments of micro-batch indices to the blocks (\texttt{IterRepetendBlocks}), following the pruned strategy outlined in Property~\ref{prop:repetend} (Line 8). Within each iteration, the solver is employed to search for the optimal schedule for the repetend (Lines 10-13).

Upon discovering a more efficient repetend, the algorithm proceeds to complete the schedule. Firstly, it identifies the warmup and cooldown blocks following \S\ref{sec:design-completion} (Lines 14-15), and then it utilizes the solver to find the time-optimal schedule for both phases (Lines 16-17). Finally, the three phases—warmup, repetend, and cooldown—are concatenated (\texttt{Concat}) to form the best schedule (Line 18). This iterative procedure continues until the lower bound is reached or the memory constraints are exceeded (Lines 19-21).

\subsection{Runtime Instantiation}
\label{sec:design-inst}

The obtained general schedule serves as a blueprint that needs to be instantiated into executable code during runtime. The schedule solely specifies the per-device execution order of blocks, without addressing the communication necessary for exchanging tensors across devices between blocks. 
% Consequently, it becomes important to generate communication plans that contain both the movement of tensors and the precise timing of these communication operations. These plans ensure the efficient coordination and synchronization of data transfers, ultimately enabling the successful execution of the scheduling plan.

% challenge: 
% 1) pair ordering: cannot mismatch send/recv order because it can cause deadlock
% 2) high efficiency: synchornized communication leading to wait 
Inserting communication operations between execution blocks presents two challenges. Firstly, it is crucial to ensure the order of communication operations, as the mis-ordering of send and receive pairs can lead to deadlocks in existing hardware.
% Maintaining the correct sequence of communication operations is essential for preventing such deadlocks and ensuring the successful exchange of tensors between devices.
Secondly, existing solutions typically rely on blocking communication, where devices must wait until the tensors are finished to move from one device to another. While this approach works well for existing schedules such as 1F1B or GPipe, it may not be suitable for all scenarios in \pn{}. In some cases, it may be infeasible to find a suitable time at which the devices involved in the communication can simultaneously finish executing their respective blocks.

% \para{Topological sort.} To address the first challenge, we employ a topological sort algorithm~\cite{toposort} to establish a consistent order for the insertion of communication operators. Specifically, we treat each pair of send and receive operators as a single operator and place them immediately after the execution block that produces the corresponding tensor. By doing so, we ensure that the communication operators are correctly positioned in relation to the data flow. Subsequently, we perform a topological sort on the schedule, taking into account the inserted communication operators. The output of this sort is a global sequential arrangement of execution blocks and communication operators that respects the partial order imposed by data dependencies and per-device block execution order. Each device then executes the assigned blocks and communication operations in accordance with this ordered sequence. As a result, we guarantee a consistent execution order for pairs of send and receive communications, avoiding the risk of mis-ordering and potential deadlocks in the system.

\para{Topological sort.} To address the first challenge, we first perform a topological sort~\cite{toposort} on the schedule. 
% The output of this sort is a global sequential arrangement of execution blocks that respects the partial order imposed by data dependencies and per-device block execution order determined by the schedule.
The output of this sort is a global sequence of execution blocks, where blocks of sharing the same starting time are placed consecutively in the sequence and follow the per-device block execution order determined by the schedule. 
% Then, we treat each pair of send and receive operators as a single operator and place them into the global sequence right after the block that produces the corresponding tensor.
Then, we treat each pair of send and receive primitives as a single operator and place them right after the time slot that produces the corresponding tensor.
During execution, each device executes the assigned blocks and communication primitives in accordance with this ordered sequence. In this way, we can guarantee a consistent execution order for pairs of send and receive communications, avoiding the risk of mis-ordering and potential deadlocks.

\begin{figure}[t!]
\centering
    \includegraphics[width=\linewidth]{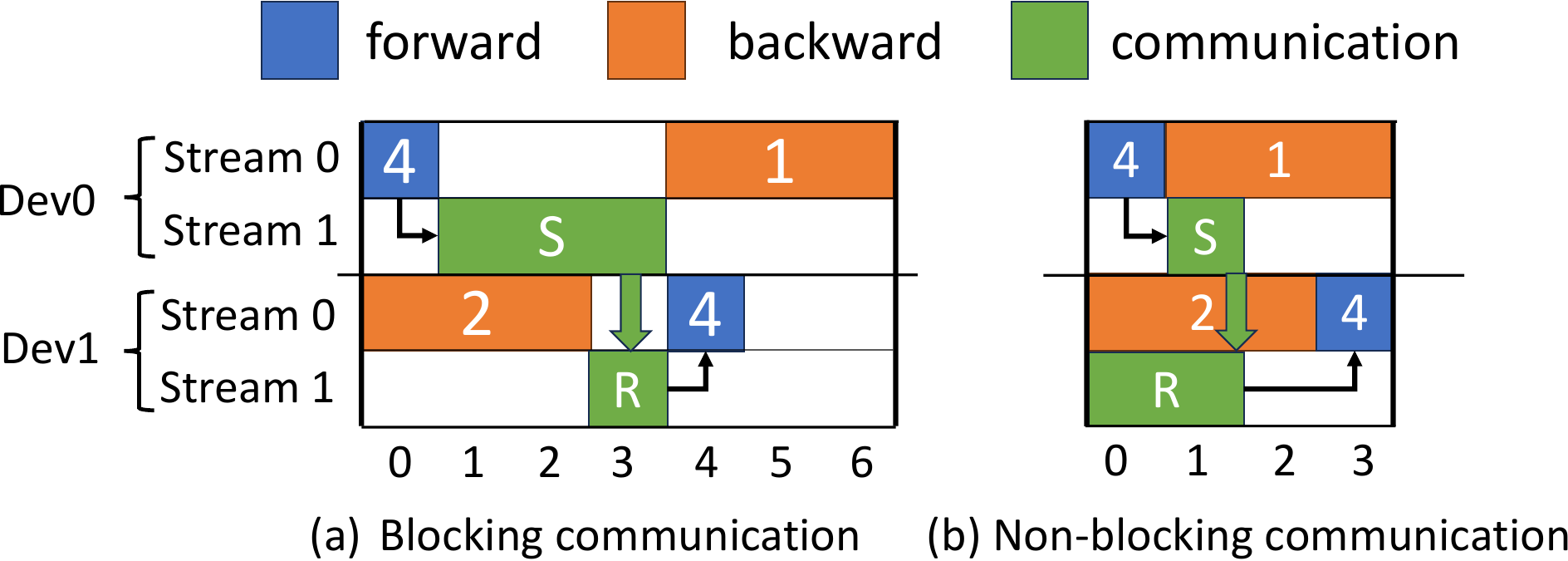}
    \caption{Non-blocking communication. The green blocks indicate the communication operators of send (S) and receive (R).}
    % \vskip -1ex
    \label{fig:async-comm}
\end{figure}

% \para{Asynchronous communication.}
\para{Non-blocking communication.} In order to enhance device utilization and mitigate potential performance degradation caused by blocking communication, we adopt a non-blocking communication pattern. Figure~\ref{fig:async-comm} demonstrates the optimization process. Suppose a tensor is generated from a forward block on device 0 and needs to be sent to device 1. As illustrated in Figure~\ref{fig:async-comm}(a), the execution of the next block has to wait until the data movement by the peer device is completed, resulting in significant idel time for device 0. By employing non-blocking communication, as shown in Figure~\ref{fig:async-comm}(b), the send and receive operations are executed concurrently with the execution blocks. As a result, the execution of blocks is not impeded by communication operations that are not dependent on them. This approach allows for more efficient utilization of device resources, as execution blocks can progress independently of communication operations that do not have direct dependency on them.

\subsection{Discussion}

\para{Optimality.} It is important to note that the problem is known to be NP-hard, which implies that finding optimal solutions is computationally infeasible in general. Consequently, \pn{} doesn't guarantee the discovery of optimal solutions. For instance, the search algorithm employed in our approach doesn't explicitly consider the opportunities arising from jointly optimizing the warmup, repetend, and cooldown phases. However, \pn{} works under the assumption that the number of micro-batches is sufficiently large, thereby ensuring that the repetend phase dominates the overall execution. While our approach may not achieve global optimality, it aims to identify efficient schedules that are effective in practice, given the constraints and computational complexity of the problem.

% \paragraph{Micro-batch execution plan.} In this paper, we assume the micro-batch execution plan as a given premise. Our primary focus lies in the search for an efficient schedule plan, rather than explicitly constructing the micro-batch execution plan. In \S\ref{sec:eval}, we present a simple policy that facilitates the construction of a micro-batch execution plan by parallelizing memory-intensive operators. However, the detailed exploration of the micro-batch execution plan are considered as future work. 
% The present work primarily concentrates on the development of a search algorithm for efficient scheduling, thereby leveraging the provided micro-batch execution plan to derive effective scheduling plans.

\section{Implementation}
\label{sec:imple}

\noindent We implemented \pn{} based on PyTorch~\cite{PyTorch}. \pn{} takes a DNN model (captured by TorchScript~\cite{torchscript}) as well as its operator placement strategy, and automatically searches for efficient schedules with device memory constraints. Then, \pn{} generates per-device PyTorch code following the searched schedule together with inserted communication primitives.

\para{Solver implementation.} Z3-solver~\cite{z3-solver} is a highly efficient Satisfiability Modulo Theories (SMT) solver known for its effectiveness in solving complex constraint satisfaction problems. We rely on Z3-solver to identify the best schedules. In our implementation, each block is associated with a variable (\texttt{z3.Int}) that represents its starting time. To encode the memory constraints, for each device, we traverse the timeline, cumulatively summing the memory cost and updating the peak memory usage by checking the existence of a block on a time slot using \texttt{z3.If}. To determine the optimal schedule, we employ a binary search approach, iteratively checking whether a given objective value can be satisfied. % \zhiqi{add explanation on memory constraint encoding.}
% we also explored Z3's optimization feature, but found it doesn't perform better than binary search.

\para{Lazy search optimization.} 
The repetend phase search involves only 1 micro-batch's blocks whereas the warmup and cooldown phases together comprise ($N_R-1$) micro-batches' blocks.
% The warmup and cooldown phases collectively comprise ($N_R-1$)-micro-batch blocks, whereas the repetend phase consists of only 1-micro-batch blocks. 
Consequently, the warmup and cooldown phases inherently require more time for searching compared to the repetend phase. To mitigate this cost, we implement a \emph{lazy search} optimization that consolidates the warmup and cooldown phase searches into a one-time process. This is achieved by checking the existence of valid schedules for the warmup and cooldown phases (\ie by replacing each time-optimal search with a single satisfiability check in Lines 16-17 of Algorithm~\ref{alg:search-overview}) once a better repetend is identified. Then after the traversal of all possible repetends, \pn{} proceeds to search for time-optimal schedules for the warmup and cooldown phases according to the best repetend. This approach significantly reduces the overall search time without changing the searched results.

% \para{Code generation.} \TODO{todo}

\para{Non-blocking communication.} Non-blocking communication is performed on a separate GPU stream, distinct from the stream used for computation. To ensure that the tensors being communicated are completed before the execution of blocks that require them, we developed a global message manager to coordinate the communication. Specifically, every non-blocking communication primitive submits the tensor instance and the corresponding communication handler to the message manager. Subsequently, prior to executing each block, the necessary tensor communications are awaited to ensure their completion. This mechanism guarantees the arrival of the required tensors before the execution of dependent blocks.

% \para{Automated micro-batch execution plan.} We firstly leverage the well formulated dynamic programming~\cite{piper} to search for balanced pipeline stages as well as nested tensor parallelism size for each stage. In order to address the varying memory consumption of different operators, we additionally employ a simple yet efficient policy by examining all operators in the model and applying tensor parallelism to divide those with higher memory consumption to all devices until the memory constraint is met. Meanwhile, it maintains well-explored execution plans for the remaining operators, such as those adopted in 1F1B~\cite{dapple} or Chimera~\cite{chimera} combining with tensor parallelism~\cite{megatron1, piper}. The resulted micro-batch execution plans are in the form of a flatten combination of tensor parallelism and traditional pipeline stage placement, which are different from conventional ways that applying tensor parallelism with a nested manner inside each pipeline stage. As the primary goal of \pn{} mainly focuses on the policy of searching efficient schedule plans given by the premise of micro-batch execution plans, we leave the exploration of more flexible micro-batch execution plans in future work.

\section{Evaluation}
\label{sec:eval}

\begin{figure*}[t!]
\centering
    \includegraphics[width=\linewidth]{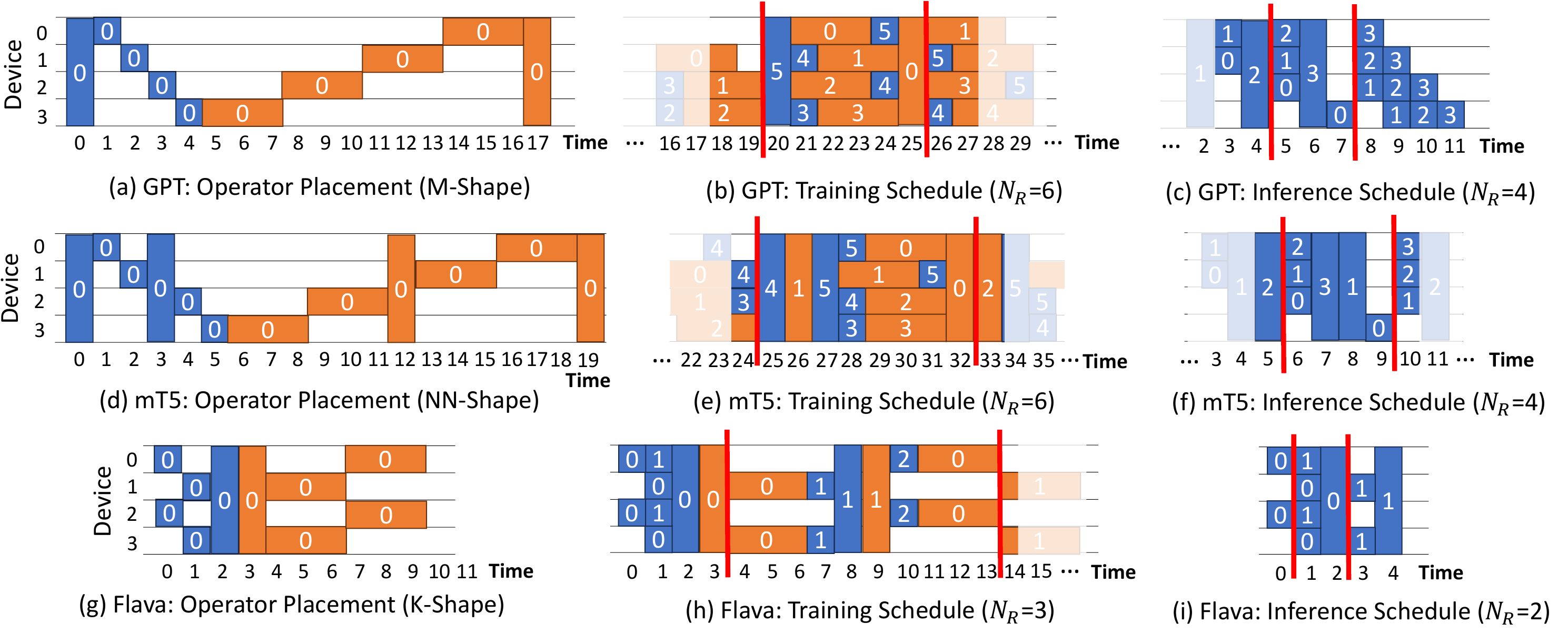}
    % \vskip -0.5ex
    \caption{Searched schedules of various models. Each row represents one model of its placement strategy, training and inference schedules from left to right. The blue and orange blocks represent forward and backward blocks, respectively. The number in each block is the micro-batch index. The execution blocks between red bars are repetends.}
    % \vskip -1ex
    \label{fig:eval-search-result}
\end{figure*}

\noindent We evaluated \pn{} using a comprehensive approach, beginning with the presentation of the searched schedules in the context of diverse operator placement strategies and models~\cite{gpt, mt5, flava}. Subsequently, we conducted ablation studies to examine the obtained results with respect to various factors. Furthermore, we compared the end-to-end training and inference performance of the schedules~\cite{dapple, chimera} predominantly employed in practice. Finally, we performed a detailed analysis of the end-to-end performance breakdown.

\subsection{Experiment Setup}

The evaluation was conducted on a 4-server cluster composed of 32 NVIDIA V100-32GB GPUs. Each server has 40-core Intel(R) Xeon(R) CPU E5-2698 v4 CPUs and 8 GPUs connected through NVLinks. Servers are interconnected via a 100 Gbps Infiniband network. The servers are equipped with NCCL v2.14, PyTorch v1.13 and Z3-solver v4.12.

\para{Models.} We evaluate \pn{} using three popular DNN models from various domains, including language and vision:
1) \textbf{GPT} model~\cite{gpt}, which is a popular model composed of homogeneous transformer layers~\cite{attention};
2) \textbf{mT5} model~\cite{mt5}, which is a multi-language encoder-decoder model, consisting of distinct types of attention blocks. During training, its encoder and decoder parts require access to a shared, large embedding table.
For GPT and mT5, we followed the recent trend~\cite{vocab500k, gpt-4} in multi-language scenarios by setting a larger embedding vocabulary size.
3) \textbf{Flava} model~\cite{flava}, which is a multi-modal model that co-considers language and vision. It incorporates separate text and vision encoders as two distinct branches, and the results of these encoders are jointly computed in a cross encoder.
Similar with Megatron-LM~\cite{megatron3} and Alpa~\cite{alpa}, we scale up the model with the total number of GPUs.

\para{Operator placement.} In \pn{}, we adopted more advanced placement strategies for each model and search schedules accordingly. These advanced strategies include distributing memory-intensive operators to all devices for all models and, for Flava, placing independent branches on distinct devices~\cite{branch-parallelism}. Figure~\ref{fig:eval-search-result}(a,d,g) illustrates the operator placement strategies for GPT, mT5, and Flava, corresponding to M-Shape, NN-Shape, and K-Shape, respectively.
Specifically, we applied full-device tensor parallelism for large embedding layers of GPT and mT5, and cross-encoder layers of Flava. For the remaining operators,
we further combined data and tensor parallelism within each block for all operator placement strategies, leveraging Piper~\cite{piper}, a state-of-the-art solution that employs dynamic programming, to search optimal configurations for these parallelisms within memory constraints.

\para{Baselines.} We compared \pn{} with three Piper-based baselines employing different schedules:
% We compare with three baselines: 
1) \textbf{1F1B}~\cite{dapple}, a popular schedule in current practice that is based on a V-Shape operator placement; 2) \textbf{Chimera(-direct)}~\cite{chimera} which leverages the X-Shape operator placement; 3) \textbf{1F1B+} which adapts the 1F1B schedule to incorporate the same advanced placement strategies used in \pn{}. To achieve this, we inserted the distributed operators closely to their neighboring operators within the 1F1B schedule, denoting it as 1F1B+.

For all baselines and \pn{}, we additionally employed the widely adopted recompute~\cite{recompute} technique to save memory for training. Following common practice~\cite{alpa, megatron2}, we enabled recompute on every transformer layer during training.

\subsection{Searching Results}

% We firstly evaluate the schedules identified by \pn{}. We take into account that the backward computation typically requires twice the amount of time as the forward computation~\cite{megatron2}. Applying recompute~\cite{recompute} further results in the backward computation taking an additional forward computation time, which triples the computational cost of the forward computation.

% We employ the recompute technique~\cite{recompute} for training, which 
The recompute technique
typically results in triple the amount of time for backward computation compared to forward computation~\cite{megatron2}.
In our evaluation of schedule efficiency, we assumed that each device has a balanced computation workload.

\begin{table}[t]
	\centering
	% \footnotesize
    \normalsize
	\begin{tabular}{c c c c c}
		\hline
		Model & 1F1B & Chimera-direct & 1F1B+ & \pn{} \\
		\hline
		GPT &   \textbf{\underline{0\%}} & 20\% & 25\% & \textbf{\underline{0\%}}  \\
        mT5 &   \textbf{\underline{0\%}} & 20\% & 20\% & \textbf{\underline{0\%}}  \\
        % Flava & \textbf{\underline{0\%}} & 20\% & 25\% & \textbf{\underline{0\%}}  \\
        Flava & \textbf{\underline{0\%}} & 20\% & $\times$ & \textbf{\underline{0\%}}  \\
		\hline
	\end{tabular}
    \caption{Bubble rate of each training schedule considering numerous micro-batches. '$\times$' indicates no straightforward adaptation of 1F1B to the given placement strategy.}
    % \vskip -2ex
    \label{tab:bubble-rate}
\end{table}

\para{Training and inference schedules.} Figure~\ref{fig:eval-search-result} displays the operator placement strategies for each model, as well as the discovered schedules for both training and inference. 
The schedules are formulated using only a small number of micro-batches (up to 6),  but they can be generalized to accommodate any large number of micro-batches. \pn{} finds schedules that can all achieve full device utilization during repetend computation for both training and inference. It's worth noting that the compact mechanism~(\S\ref{sec:design-construct}) of repetends enables \pn{} to discover more flexible repetends, such as those shown in Figure~\ref{fig:eval-search-result}(c, f, h), which may appear to have device idle time but can actually be reduced during runtime. Interestingly, in most cases, by simply excluding the execution of backward blocks within the training schedules, we find that the training and inference schedules can share the same execution schedule during the final runtime, despite having different representations of repetends. This indicates that the inference schedules can be easily obtained by selectively excluding the execution of backward blocks and their corresponding communication operations from their training schedules.

\para{Schedule efficiency.} We evaluated the schedule efficiency using bubble rate~\cite{megatron1}, a common metric calculated by the occupation of device idle time during the entire execution. Table~\ref{tab:bubble-rate} presents a comparison between \pn{} and three other baselines. While \pn{} shares the same placement strategy as 1F1B+, it can search for more efficient schedules. Although 1F1B can achieve zero bubble rate under its V-Shape placement strategy, it may suffer from inefficiency due to workload imbalances during real runtime~(\S\ref{sec:eval-runtime}).

\begin{figure}[t]
    \centering
    % \vskip 1ex
    % \includegraphics[width=0.99\linewidth]{figures/eval-search-time.pdf}
    \includegraphics[width=0.99\linewidth]{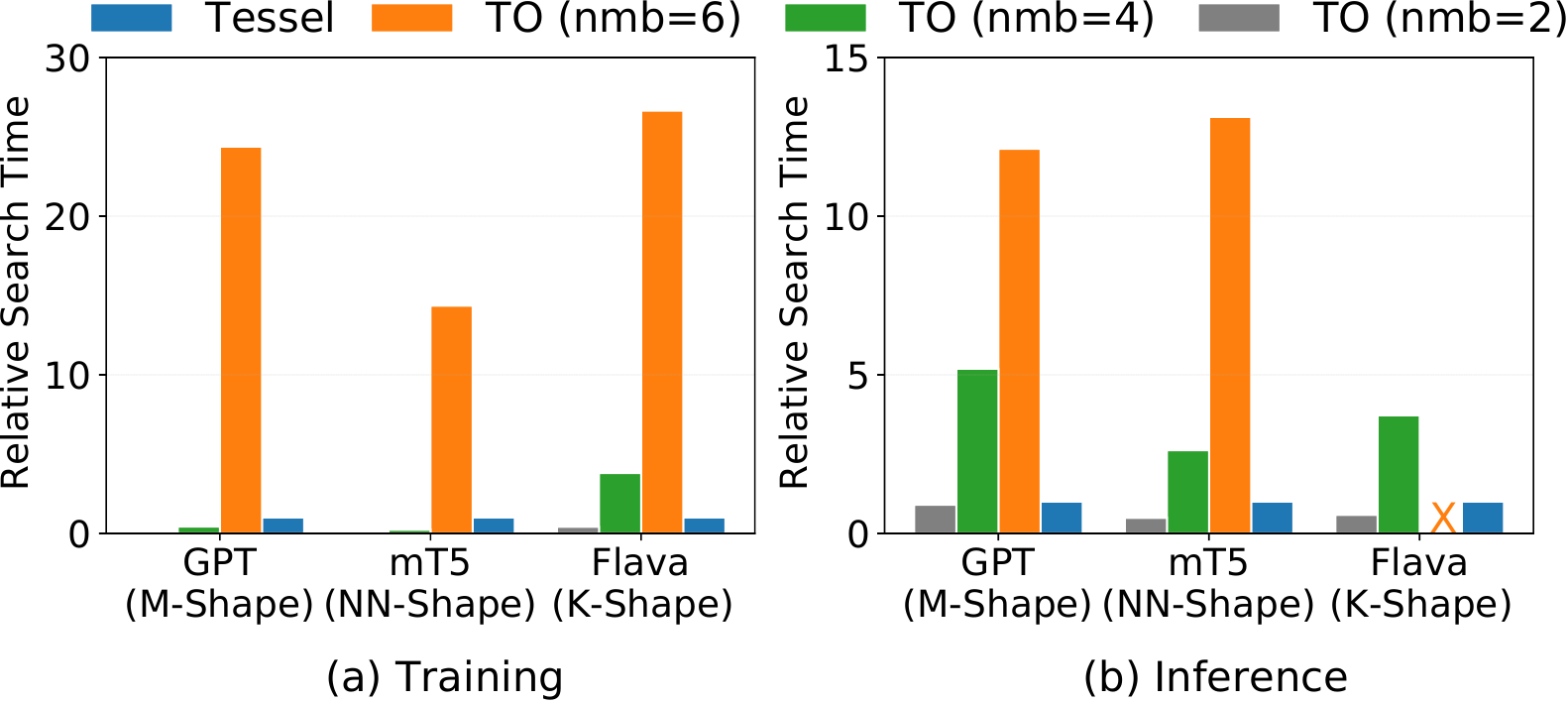}
    \caption{Search cost of TO normalized by \pn{} time. `$\times$' denotes value exceeding 10k.}
    \label{fig:eval-search-cost}
\end{figure}

\para{Search cost.} To demonstrate \pn{}'s efficient search process over a large schedule space, we compared the search time with the time-optimal search~(\S\ref{sec:motivation}) approach, referred to as TO. Since it is impractical to use TO to search schedules with a large number of micro-batches, we ran TO with different small numbers of micro-batches (nmb) and compared the search time with \pn{}. Figure~\ref{fig:eval-search-cost} illustrates the results, which show that \pn{} significantly reduces search time compared to the baseline solution, indicating its efficiency. 
While \pn{} does not guarantee optimal search results, we observed that when we extended the searched schedules of \pn{} to the same number of micro-batches searched by TO, the resulting schedules exhibited the same bubble rates, \eg both TO and \pn{} reached 20\% of the bubble rate of NN-Shape for 6 micro-batches, considering both warmup and cooldown phases. This observation suggests that \pn{} is able to achieve close-optimal solutions in the majority of cases.

\begin{figure}[t]
    \centering
    % \vskip 1ex
    \includegraphics[width=0.99\linewidth]{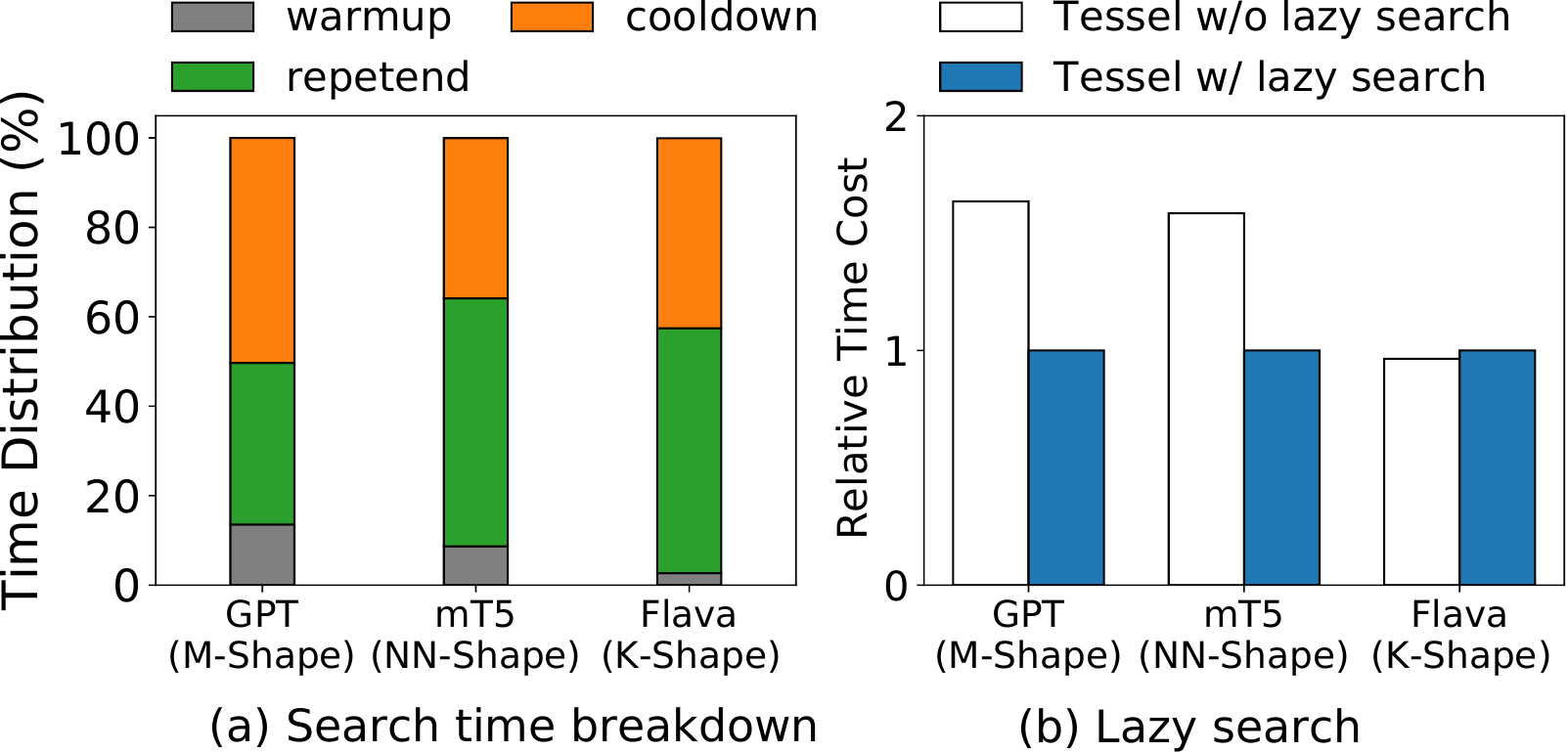}
    \caption{Search time breakdown with lazy search optimization.}
    \label{fig:eval-search-breakdown}
\end{figure}

\para{Search time breakdown.} 
We further conducted a breakdown of the search time to study the cost distribution of different phases in \pn{}. 
On average, \pn{} spends 147.3 seconds to finish the search.
Figure~\ref{fig:eval-search-breakdown} (a) shows the time distribution.
% A larger number of blocks in the search problem corresponds to a larger search space and consequently more time spent. 
% Following this, 
We made several observations:
1) Teh cooldown phase tends to require more search time than the warmup phase due to the larger number of blocks in the former phase;
2) Despite the warmup and cooldown phases having more blocks than the repetend phase, their search time remains comparable to that of the repetend phase. This is primarily attributed to the application of lazy search optimization, which streamlines the search into an one-time process after iterating through all repetends. Figure~\ref{fig:eval-search-breakdown}(b) supports this observation by comparing the relative search time costs with and without lazy search optimization;
3) The search time of the repetend phase varies depending on the operator placement and is influenced by the traversal order of all the repetends, potentially leading to early termination when a zero-bubble repetend is discovered.

\subsection{Searching Ablation Study}
\label{sec:eval-ablation}

We further conducted ablation studies to evaluate the factors that influenced the search results. We evaluated two factors: 1) the maximal number of micro-batches ($N_R$) involved in the repetend construction; 2) the memory capacity ($M$).

\begin{figure}[t]
    \centering
    % \vskip 1ex
    \includegraphics[width=0.95\linewidth]{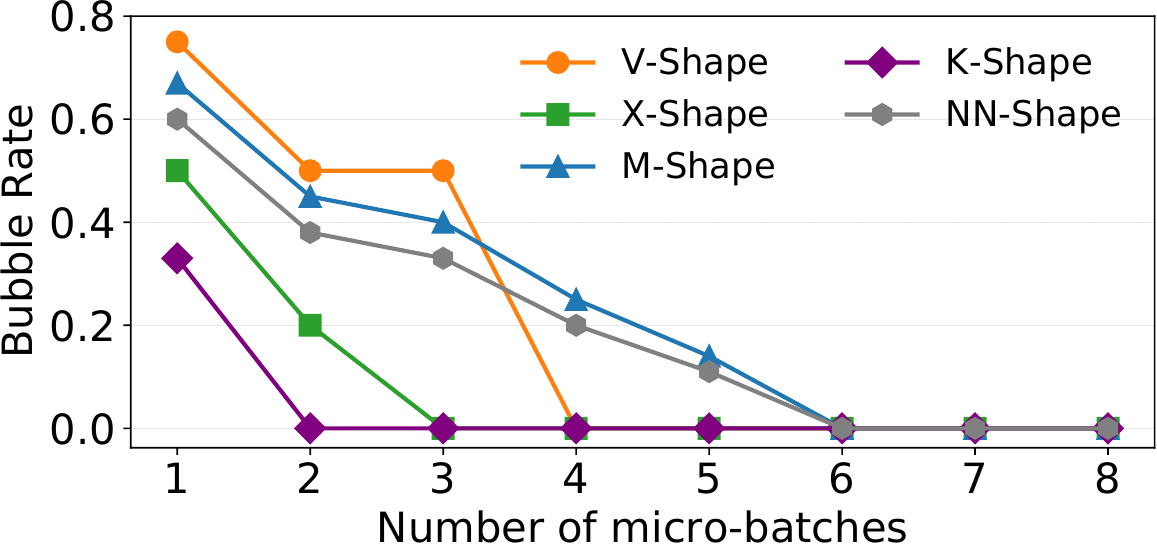}
    \caption{Bubble rate in terms of micro-batch numbers ($N_R$).}
    \label{fig:eval-ablation-nmb}
\end{figure}

\para{Bubble rate and $N_R$.} The bubble rate can be influenced by the number of micro-batches ($N_R$) involved in repetend construction. Figure~\ref{fig:eval-ablation-nmb} shows the bubble rate as $N_R$ increases for various operator placement strategies when the memory capacity is not constrained. Interestingly, we observed that all schedules can achieve a zero bubble rate when a sufficiently large number of micro-batches is used. However, it is worth noting that the starting number of micro-batches required to achieve a zero bubble rate differs across different operator placement strategies. For instance, the V-Shape placement requires a minimum of 4 micro-batches, which corresponds to the number of devices, while the NN-Shape and M-Shape placements requires at least 6 micro-batches. This suggests the necessity to search for a feasible number of micro-batches to achieve low device idle time.

\begin{table*}[t]
	\centering
	\small
	\begin{tabular}{c c c c c c c}
		\hline
		\textbf{Model} & \textbf{Parameters} & \textbf{Layer} & \textbf{Hidden Size} & \textbf{Head Number} & \textbf{Vocabulary Size}  \\
		\hline
		GPT~\cite{gpt-3} & \{11B, 24B, 47B, 77B\} & \{32, 40, 48, 80\} & \{4096, 6144, 8192, 8192\} & \{32, 48, 64, 64\} & \{1M, 1M, 1M, 1.5M\} \\
        mT5~\cite{mt5} & \{1.8B, 9.5B, 43B, 88B\} & \{48, 48, 64, 80\} & \{1024, 3072, 6144, 8192\} & \{16, 24, 48, 64\} & \{512K, 1M, 1.5M, 1.5M\} \\ 
		\hline
	\end{tabular}
	% \vskip 2ex
	\caption{Model architecture with increasing number of GPUs. K: thousand. M: million. B: billion.}
	% \vskip -3ex
	\label{tab:model}
\end{table*}

\begin{figure}[t]
    \centering
    % \vskip 1ex
    % \includegraphics[width=0.85\linewidth]{figures/eval-ablation-memory.pdf}
    \includegraphics[width=0.95\linewidth]{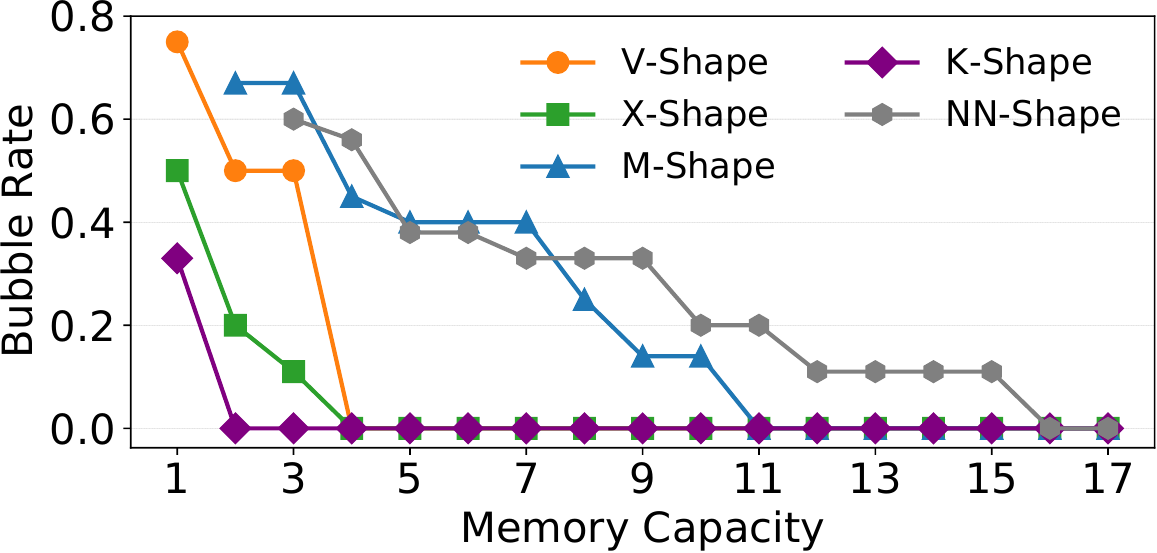}
    \caption{Bubble rate in terms of memory capacity ($M$).}
    \label{fig:eval-ablation-memory}
\end{figure}

\para{Bubble rate and $M$.} To further investigate the impact of memory capacity ($M$) on the bubble rate, we conducted an ablation study by increasing memory capacity. In this study, we maintained the same starting $N_R$ that allowed us to achieve a zero bubble rate in Figure~\ref{fig:eval-ablation-nmb}. 
For simplicity, we 
considered the memory consumption of each forward and backward block as 1 and -1, respectively. Figure~\ref{fig:eval-ablation-memory} presents the bubble rate in relation to memory capacity. We observed that a lower memory capacity corresponds to a larger bubble rate. This is because it requires fewer forward execution blocks to be executed before the first backward execution block, thus filtering out schedules that could have executed them earlier for better device utilization. Similarly, as the memory capacity becomes sufficiently large, all schedules can ultimately achieve a zero bubble rate.

We noticed a strong correlation between $N_R$ and $M$ (Figures~\ref{fig:eval-ablation-nmb} and ~\ref{fig:eval-ablation-memory}).  
We observed that the trend of the bubble rate tends to be similar 
% when the maximal number of in-flight micro-batches and the maximal number of micro-batches involved in repetend construction are the same.
when increasing $N_R$ and $M$.
Intuitively, a larger $N_R$ exposes a broader space for scheduling, allowing more forward execution blocks to be executed ahead of time and thereby improving device utilization. Similar results are observed for higher memory capacity.

\subsection{End-to-end Runtime Performance}
\label{sec:eval-runtime}

% We implement the above baselines on our system, with each reaching to the similar bubble rate of the theoretical estimation (\S\ref{sec:eval-performance-breackdown}).

% In addition to the pipeline parallelism algorithms, we incorporate tensor parallelism~\cite{megatron1} for each pipeline stage when memory capacity is insufficient. We leverage dynamic programming algorithm proposed in Piper~\cite{piper} to determine the tensor parallelism size for each pipeline stage, as well as the operator assignment for each stage. Due to the co-location of different stages on same devices in Chimera, we enforce a constraint on the tensor parallelism size to maintain consistency across all stages. Furthermore, we apply the recompute~\cite{recompute} mechanism to each layer in order to save memory.

We evaluated the training performance of GPT and mT5, and the inference performance of Flava. For other scenarios, \eg inference of GPT, \pn{} demonstrates comparable performance to the baselines. We set the global size to 128 during training and tuned the micro-batch size to reach the best performance. Table~\ref{tab:model} illustrates the model configurations used during the evaluation as we increased the number of GPUs. Similar to Alpa~\cite{alpa}, we used aggregated Peta Floating-point Operations Per Second (PFLOPS) as the performance metric during training.

\begin{figure}[t]
    \centering
    \includegraphics[width=0.95\linewidth]{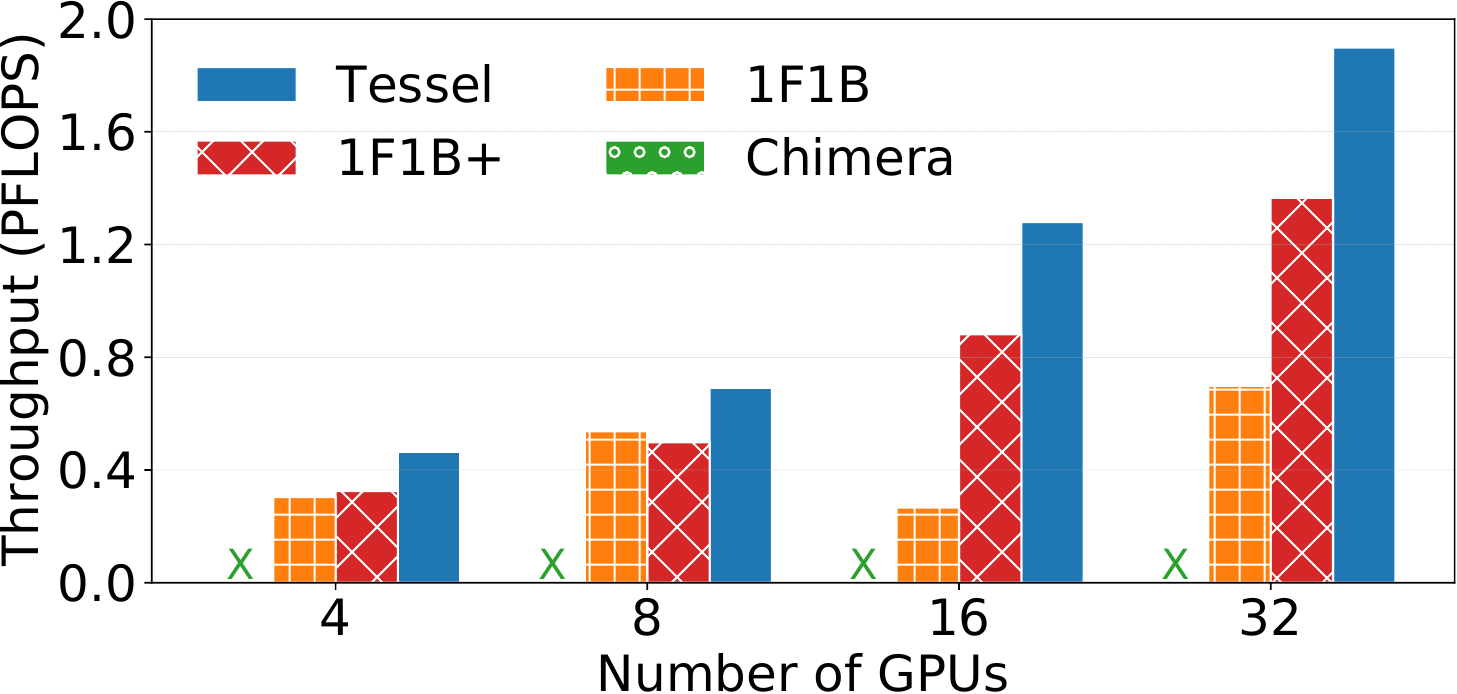}
    \caption{GPT end-to-end training throughput. ($\times$: failure due to out of memory)}
    % \vskip -1ex
    \label{fig:eval-gpt-e2e}
\end{figure}

% \para{GPT training results.} Figure~\ref{fig:eval-gpt-e2e} illustrates the performance of \pn{} and baselines on GPT training. Overall, \pn{} reaches up to 5.8$\times$ (16-GPU) and 1.4$\times$ speedup comparing with 1F1B and 1F1B+, respectively. Chimera fails to run in this scenario due to out-of-memory issues caused by its placement strategy by co-locating parameters of multiple stages within a single GPU, exacerbating the memory bottleneck. 1F1B requires cross-node tensor parallelism to distributed the large embedding layer due to memory constraints, which suffers from heavy communication overhead as the co-located transformer layers with the embedding layer also requires to communicate tensors across nodes. 1F1B+ instead employs a better placement strategy which only requires embedding layer to perform cross-node communication, which outperforms 1F1B when the GPU number is large. However, it still suffers from inefficiency due to data dependency. \pn{} leverages the same placement strategy with 1F1B+ but is able to search for a more efficient schedule, hence having better performance.

\para{GPT training results.} Figure~\ref{fig:eval-gpt-e2e} illustrates the performance comparison between \pn{} and the baselines during GPT training. \pn{} achieved up to 4.8$\times$ (16-GPU) and 1.4$\times$ speedup compared to 1F1B and 1F1B+, respectively. Chimera failed to run in this scenario due to out-of-memory issues caused by its placement strategy, which co-located parameters of multiple stages within a single GPU, exacerbating the memory bottleneck. In multi-server cases, 1F1B 
% requires
encountered out-of-memory issues, with intra-server tensor parallelism necessitating the application of cross-server tensor parallelism to distribute the large embedding layer. This lead to heavy communication overhead. 1F1B+ instead adopted the M-Shape placement strategy that only required the embedding layer to perform cross-server communication, improving performance by saving communication costs. However, 1F1B+ still suffered from inefficiency due to data dependency. \pn{} adopted the same placement strategy as 1F1B+ and outperformed it by searching for a more efficient schedule.

% \para{GPT training results.} Figure~\ref{fig:eval-gpt-e2e} illustrates the performance of \pn{} and baselines on GPT training. When using a small number of GPUs, 1F1B and GPipe encounter high memory usage on the pipeline stage with the large embedding layer. This leads to imbalanced computation across stages, resulting in idle time on GPUs during execution. In contrast, \pn{} parallelizes the large embedding weight across all devices, alleviating the memory bottleneck and achieving a balanced computation workload for each device, which improves performance. With a larger number of GPUs (more than 8), 1F1B and GPipe can achieve balanced pipeline stages. However, they require cross-node tensor parallelism to support the large embedding layer and the co-located transformer layers, incurring heavy communication costs. \pn{}, on the other hand, only parallelizes the embedding layer while keeping the remaining transformer layers parallelized within the node, resulting in little communication overhead. As for Chimera, it fails to run in this scenario due to out-of-memory issues caused by its strategy of placing parameters of multiple stages within a single GPU, exacerbating the memory bottleneck.

\begin{figure}[t]
    \centering
    \includegraphics[width=0.9\linewidth]{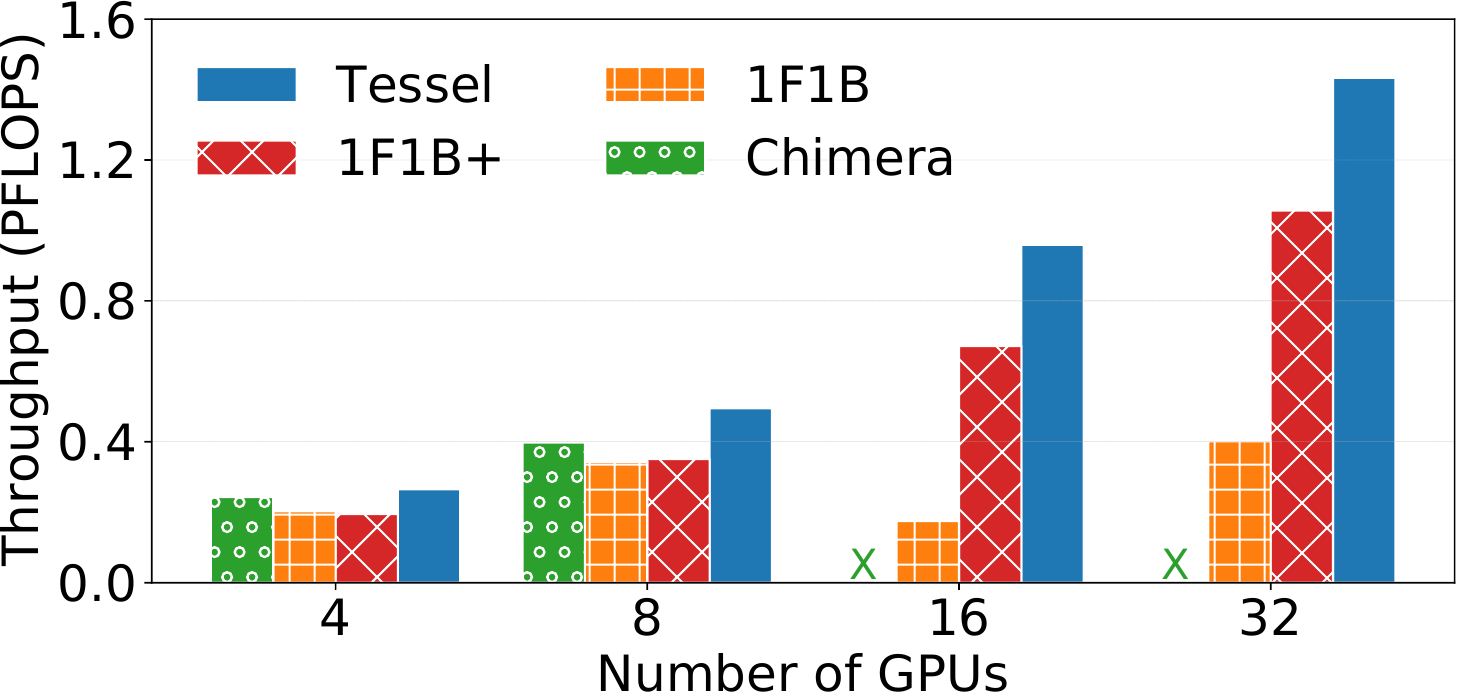}
    \caption{mT5 end-to-end training throughput. ($\times$: failure due to out of memory)}
    \label{fig:eval-mt5-e2e}
\end{figure}

\para{mT5 training results.} Figure~\ref{fig:eval-mt5-e2e} compares the performance of \pn{} with the baselines on mT5 training. Overall, \pn{} achieves up to 5.5$\times$ and 1.4$\times$ performance speedup compared to existing schedules and 1F1B+, respectively. Within one server, Chimera slightly outperformed 1F1B+, due to the lower bubble rate indicated in Table~\ref{tab:bubble-rate}. When scaled to multiple servers, 1F1B+ outperformed 1F1B for similar reasons as observed in the GPT results. \pn{} still maintained its superior performance over all baselines by identifying the zero-bubble schedule for the advanced placement.

% \para{mT5 training results.} Figure~\ref{fig:eval-mt5-e2e} compares the iteration time of \pn{} with baselines on mT5 training. \pn{} achieves up to 5.3$\times$ (16-GPU) performance speedup over the best of existing baselines. Similar with the results of GPT, when using smaller models on less GPUs, \pn{} shows similar results with all baselines. However, witch larger models on more GPUs, \pn{} achieves 5.3$\times$ and 4.3$\times$ performance speedup over 1F1B on 16 and 32 GPUs, respectively. This is because the more efficient memory utilization that enables \pn{} to use intra-node tensor parallelism for encoder and decoder layers. GPipe can easily run out of memory, which is because all the output tensors at the re-compute layer boundaries are required to be kept, which consume lots of memory.

\begin{figure}[t!]
    \centering
    \includegraphics[width=0.99\linewidth]{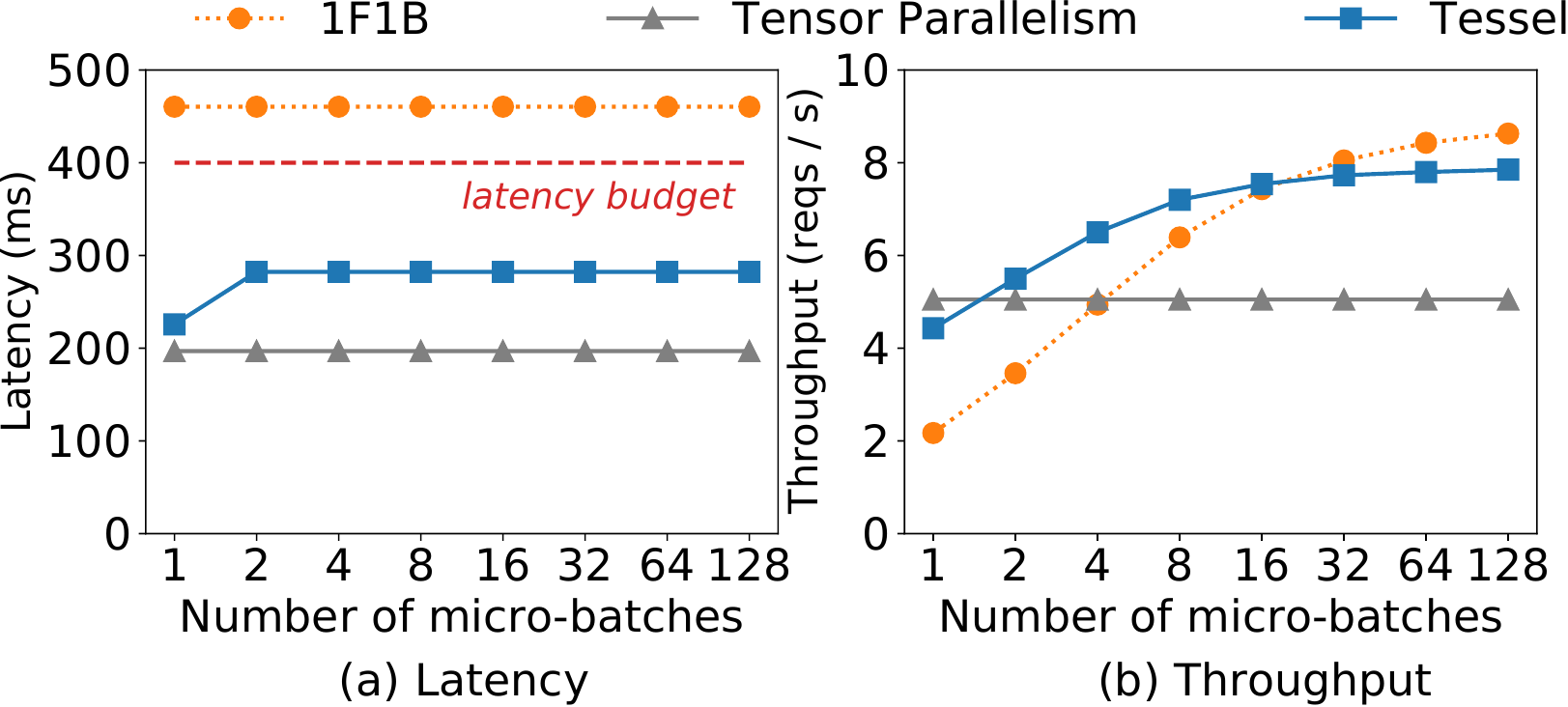}
    \caption{Inference latency and throughput of Flava (24 layers, 4096 hidden size with 32 heads) on 4 GPUs.}
    \label{fig:eval-flava-e2e}
\end{figure}

\para{Flava inference results.}
% \pn{} excels not only in generating efficient training schedules but also in searching efficient schedules for inference. 
Inference workloads often come with latency budget requirements, with 400 ms being a recommended setting based on previous studies~\cite{sla-400ms, sla-400ms-2}. While meeting the latency budget is crucial, service providers can further benefit from increasing throughput to save costs. Therefore, pipeline parallelism can be an effective choice to further enhance throughput while adhering to latency constraints.
For the inference evaluation, we used the Flava model on 4 GPUs. 
Since there is no straightforward adaptation of 1F1B to the K-Shape, we solely compared the conventional 1F1B and tensor parallelism~\cite{megatron1} with \pn{}.
% As Chimera's latency matches 1F1B's, we omit it from the comparison.

% Figure~\ref{fig:eval-flava-e2e} illustrates the inference performance trade-offs for various micro-batch sizes.
% Overall, \pn{} unleashes the benefit of the advanced operator placement to reduce latency by 38.7\% for executing one micro-batch, while achieving similar throughput when compared with 1F1B schedule.
% \gb{Is this because of placement? Should our schedule be better than 1F1B in the same placement?}.
% The 
Figure~\ref{fig:eval-flava-e2e} illustrates the latency-throughput trade-offs for various micro-batch sizes.
Notably, 1F1B focuses on optimizing throughput, while tensor parallelism prioritizes latency. In contrast, \pn{} demonstrates a more balanced trade-off by slightly increasing latency within the budget while significantly improving throughput, resulting in 1.5$\times$ throughput speedup compared to tensor parallelism.
1F1B demostrates high throughput only for large micro-batch sizes, but it falls short in optimizing latency, consistently failing to meet the latency budget.
% \pn{} Both \pn{} and tensor parallelism meet the latency requirement while 1F1B does not. Furthermore, \pn{} achieves 1.5$\times$ higher throughput than tensor parallelism due to more efficient computation.
In \pn{} with 1F1B, the
latency speedup is attributed to the concurrent execution of independent branches on multiple devices, whereas 1F1B can only schedule the branches in sequential execution order.
For a large number of micro-batches (\eg 128), the throughput of \pn{} is slightly lower than 1F1B due to kernel inefficiency when applying tensor parallelism on cross-encoder model parts. However, when the number of micro-batches is small, \pn{} can achieve up to 2.0$\times$ throughput speedup over 1F1B. This boost is attributed to the lower latency of a single micro-batch execution, resulting in an overall shorter time to execute a small number of micro-batches.
For tensor parallelism, 1F1B partitions operators into smaller ones that may not fully saturate the GPU during computation. In contrast, \pn{} places independent branches on different devices without partitioning operators, preserving computation efficiency and improving throughput.
% further improved

% \para{Flava inference results.} \pn{} not only excels in generating efficient training schedules but also demonstrates its prowess in searching optimized schedules for inference. For the inference evaluation, we use Flava on 4 GPUs. Since the K-Shape placement significantly differs from V-Shape, adapting 1F1B schedule to fit K-Shape proves challenging. Therefore, we solely compare the conventional 1F1B with \pn{}. Figure~\ref{fig:eval-flava-e2e} illustrates the inference latency and throughput for various micro-batch sizes. Overall, \pn{} leverages the benefits of advanced operator placement to reduce latency by 38.7\% for executing one micro-batch, while achieving similar throughput when compared with 1F1B schedule. The latency speedup is attributed to the concurrent execution of independent branches on multiple devices, whereas 1F1B schedule can only schedule the branches in sequential execution order. For very large numbers of micro-batches (\eg 128), \pn{}'s throughput is slightly lower than 1F1B due to kernel inefficiency when applying tensor parallelism on cross-encoder model parts. However, when the number of micro-batches is small, \pn{} achieves up to 2.0$\times$ throughput speedup over 1F1B. This boost is attributed to the lower latency of one micro-batch execution, resulting in an overall shorter time for executing a small number of micro-batches.

\subsection{Runtime Performance Analysis}
\label{sec:eval-performance-breackdown}

\begin{figure}[t]
    \centering
    % \vskip 2ex
    \includegraphics[width=0.99\linewidth]{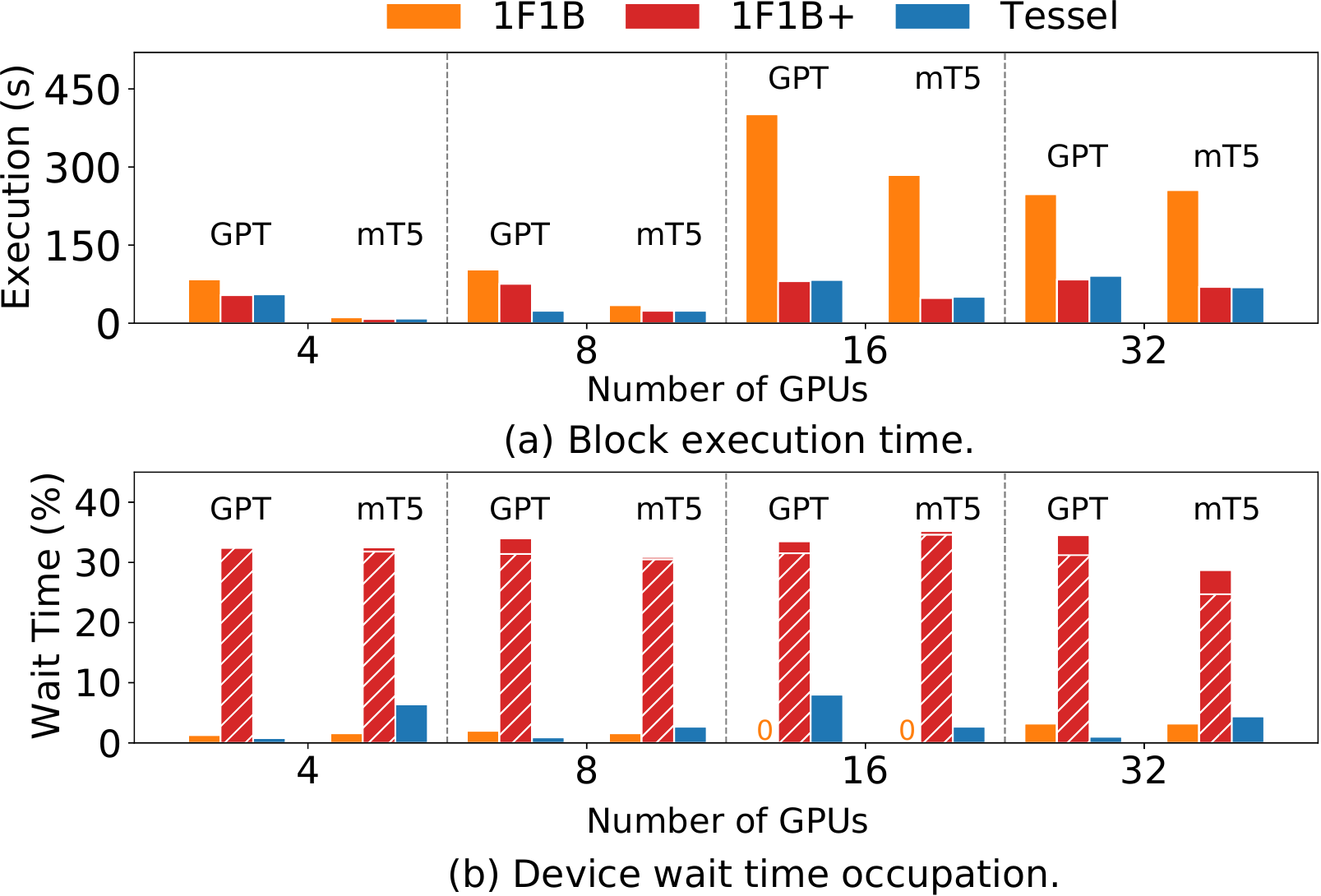}
    \caption{Runtime performance breakdown of (a) block execution time and (b) device wait occupation. The slashed region in (b) shows the theoretical estimation of wait time occupation.}
    % \vskip -1ex
    \label{fig:eval-runtime-wait-bias}
\end{figure}

\begin{figure}[t]
    \centering
    \includegraphics[width=0.99\linewidth]{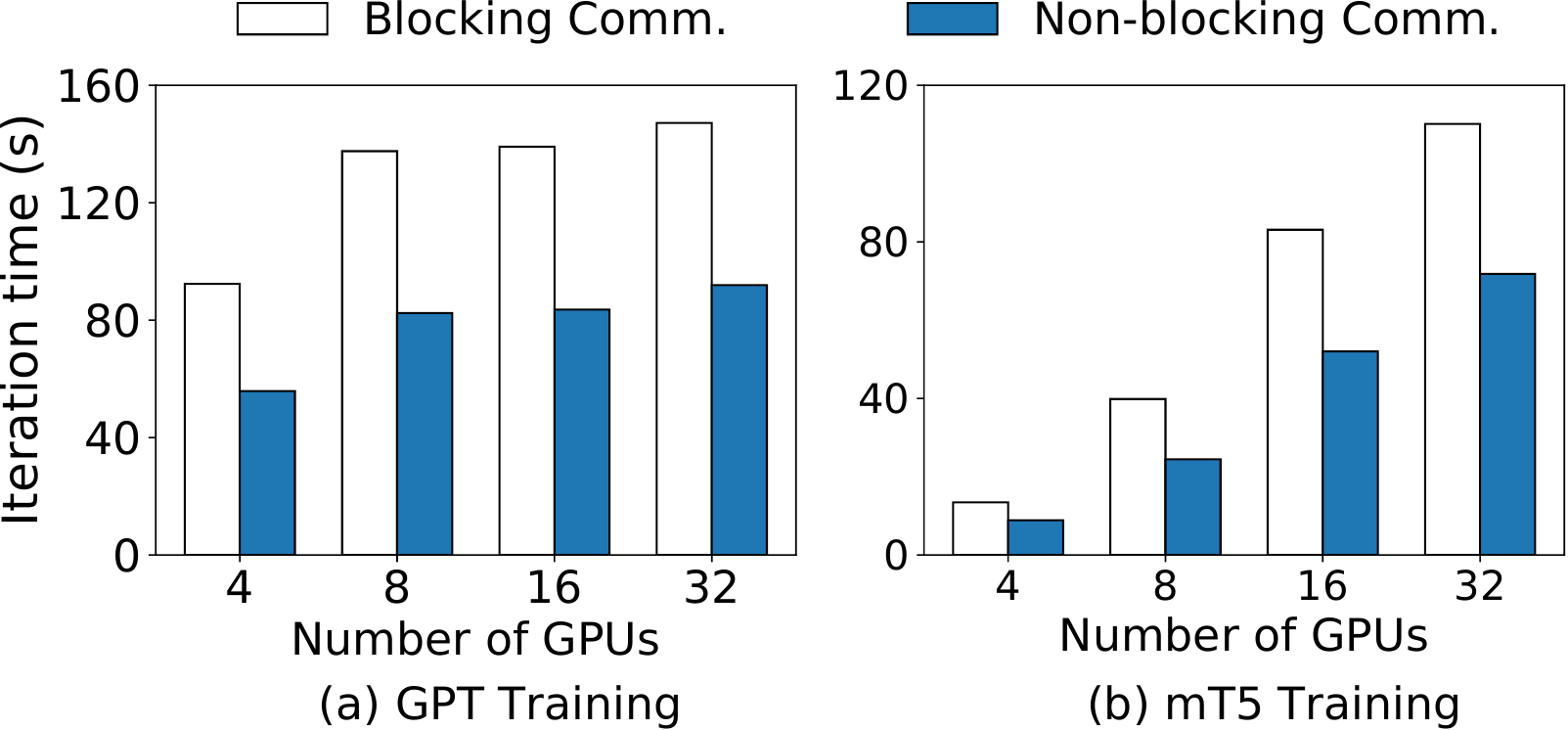}
    % \vskip -0.5ex
    \caption{End-to-end training time of blocking and non-blocking communication of GPT (M-Shape) and mT5 (NN-Shape).}
    % \vskip -3ex
    \label{fig:eval-async-comm}
\end{figure}

\para{Performance breakdown.} Runtime execution can be divided into two components: block execution time and device waiting time. Device waiting time refers to the time span between the execution of neighboring blocks, including data communication between blocks and device idle time caused by data dependency. To accurately determine device waiting time in the schedules, we profiled the runtime at the slowest stage. Figure~\ref{fig:eval-runtime-wait-bias} displays (a) block execution time and (b) wait time occupation in relation to end-to-end training performance. Overall, wait time occupation remains below 6\% overhead of the theoretical estimation (slashed region in Figure~\ref{fig:eval-runtime-wait-bias}(b)), showcasing \pn{}'s efficient runtime implementation. Both 1F1B and \pn{} have theoretically zero bubble rate and achieve low device wait time occupation. However, \pn{} outperforms 1F1B significantly due to its more balanced block execution workloads across devices. For instance, in GPT training with 16 GPUs, 1F1B requires almost 400 seconds for block computation on the slowest device, while \pn{} only requires around 100 seconds. Comparing \pn{} with 1F1B+, both employ the same placement strategy, resulting in similar block execution costs. However, \pn{} outperforms 1F1B+ by searching for more efficient schedules that significantly reduce device wait time during runtime.

% \para{Asynchronous communication.}
\para{Non-blocking communication.} \pn{} adopts non-blocking communication to overlap the communication with the computation, resulting in improved device utilization. In Figure~\ref{fig:eval-async-comm}(a) and (b), we show the end-to-end training performance comparison of \pn{} with both blocking and non-blocking communication on GPT and mT5, respectively. By utilizing non-blocking communication, \pn{} achieves up to 1.9$\times$ speedup. We do not show the performance of schedules that are based on V-Shape, X-Shape, and K-Shape, as their communication can occasionally happen at the same time between execution blocks, resulting in similar performance.

\section{Related Work}

% scheduling algorithms
% 1) improve efficiency in stage content
% 2) improve efficiency using an improved schedule
\para{Schedules.} Existing research has primarily focused on 
generally applicable efficient schedules.
% optimizing schedules to enhance device utilization. 
Notably, 1F1B~\cite{dapple, megatron2, pipedream} and GPipe~\cite{gpipe} are popular schedules widely applied in practical settings. Many works have built upon 1F1B to further enhance efficiency by incorporating additional techniques~\cite{varuna, alpa-schedule, hetpipe}. For example, Varuna~\cite{varuna} considers schedules with recompute by eliminating recompute at the last pipeline stage. Some schedules also take memory constraints into account~\cite{gems, chimera}. For instance, Gems~\cite{gems} proposes a memory-efficient schedule by considering two model replicas. 
These handcrafted schedules can also be searched automatically by using \pn{}. Furthermore, \pn{} can also accommodate the operator placements that are beyond the support of such predefined schedules. 
% These schedules are manually optimized and can be automatically searched by \pn{}. 

\para{Software pipeline optimizations.} 
In the domain of parallel computing, similar pipeline problems~\cite{allan1995software} revolve around efficiently parallelizing instructions within loop code. These optimizations also involve identifying efficient kernels (\ie similar to repetend in \pn{}) by conducting a thorough analysis of data dependencies among instructions and capitalizing on hardware capabilities~\cite{chen1994profile, rau1992register, kim2012improving, urpr}. For example, URPR~\cite{urpr} unrolls the loop by several iterations and re-orders instructions to identify kernels. In the context of DNN execution, \pn{} tackles a unique pipeline problem characterized by explicit data dependencies among blocks, variable computation cost of blocks and additional constraints such as memory capacity.

\para{Parallelization techniques.} In addition to the schedules related to operator placement strategies, there are other parallelization~\cite{data-parallel, flexflow, tofu, megatron1, recompute, deepspeed, zero, swapadvisor} techniques that can help improve performance. For example, tensor parallelism~\cite{flexflow, tofu, megatron1} enables the partitioning of a single operator across multiple devices for concurrent execution, and DeepSpeed~\cite{deepspeed} leverages ZeRO~\cite{zero, zero-offload} to offload tensors to CPU or other devices. These techniques are complementary to the schedule and can be viewed as operator placement for \pn{} to generate corresponding schedules.
% which can be jointly considered to further improve performance. % \zhiqi{maybe an example for demonstration.}

% automated parallelization
\para{Automated parallelization.} The automation of distributed DNN training and inference is important for large and diverse models. Existing works~\cite{alpa, piper, gspmd} mainly consider combining tensor parallelism with pipeline parallelism and search for feasible configurations of each parallelism. For example, Alpa~\cite{alpa} uses an ILP solver for tensor parallelism and dynamic programming for pipeline parallelism. Works like theses are designed to adpot a predefined schedule like 1F1B.
% pipeline parallelism.
Complementing \pn{}, these search algorithms can further extend their various operator placement strategies using \pn{}'s schedule search for better performance.
% to search efficient schedules under various operator placement strategies. 

\section{Conclusion}

\noindent 
\pn{} is an efficient and automated DNN schedule-searching system that can accommodate diverse operator placement strategies. 
Based on the observation of repetends in DNN schedules, \pn{} greatly reduces the schedule search space while delivering high performance results. 
We believe that \pn{} can help parallel DNN training and inference systems to better exploit their performance. 
More importantly, we hope that the DNN schedule properties that \pn{} has revealed can inspire more relevant research in the future.
% The capability of \pn{} to search for schedules paves the way for future automation in more flexible search space, offering promising opportunities to better exploit performance.
% In conclusion, \pn{} is an automated schedule-searching system that efficiently searches for highly effective DNN schedules accommodating to various operator placement strategies and implements them in real runtime execution. By the key observation of repetend in schedules, \pn{} introduces a two-phase searching algorithm that greatly reduces the search space while delivering high performance. The capability of \pn{} to search for schedules paves the way for future automation in more flexible search space, offering promising opportunities to better exploit performance.

\section{Acknowledgments}

\noindent
We thank all the anonymous reviewers for their insightful comments during the paper reviewing period. This work is supported in part by the National Natural Science Foundation of China under Grant No. 62141216, 62172382, and 61832011. Cheng Li and Youshan Miao are the corresponding authors.

%%%%%%% -- PAPER CONTENT ENDS -- %%%%%%%%

%%%%%%%%% -- BIB STYLE AND FILE -- %%%%%%%%
\bibliographystyle{IEEEtranS}
\bibliography{reference}

% Generated by IEEEtranS.bst, version: 1.13 (2008/09/30)
\begin{thebibliography}{10}
\providecommand{\url}[1]{#1}
\csname url@samestyle\endcsname
\providecommand{\newblock}{\relax}
\providecommand{\bibinfo}[2]{#2}
\providecommand{\BIBentrySTDinterwordspacing}{\spaceskip=0pt\relax}
\providecommand{\BIBentryALTinterwordstretchfactor}{4}
\providecommand{\BIBentryALTinterwordspacing}{\spaceskip=\fontdimen2\font plus
\BIBentryALTinterwordstretchfactor\fontdimen3\font minus \fontdimen4\font\relax}
\providecommand{\BIBforeignlanguage}[2]{{%
\expandafter\ifx\csname l@#1\endcsname\relax
\typeout{** WARNING: IEEEtranS.bst: No hyphenation pattern has been}%
\typeout{** loaded for the language `#1'. Using the pattern for}%
\typeout{** the default language instead.}%
\else
\language=\csname l@#1\endcsname
\fi
#2}}
\providecommand{\BIBdecl}{\relax}
\BIBdecl

\bibitem{allan1995software}
V.~H. Allan, R.~B. Jones, R.~M. Lee, and S.~J. Allan, ``Software pipelining,'' \emph{ACM Computing Surveys (CSUR)}, 1995.

\bibitem{varuna}
S.~Athlur, N.~Saran, M.~Sivathanu, R.~Ramjee, and N.~Kwatra, ``Varuna: scalable, low-cost training of massive deep learning models,'' in \emph{Proceedings of the Seventeenth European Conference on Computer Systems}, 2022, pp. 472--487.

\bibitem{gpt-3}
T.~Brown, B.~Mann, N.~Ryder, M.~Subbiah, J.~D. Kaplan, P.~Dhariwal, A.~Neelakantan, P.~Shyam, G.~Sastry, A.~Askell \emph{et~al.}, ``Language models are few-shot learners,'' \emph{Advances in neural information processing systems}, 2020.

\bibitem{recompute}
T.~Chen, B.~Xu, C.~Zhang, and C.~Guestrin, ``Training deep nets with sublinear memory cost,'' \emph{arXiv preprint arXiv:1604.06174}, 2016.

\bibitem{chen1994profile}
W.~Y. Chen, S.~A. Mahlke, N.~J. Warter, S.~Anik, and W.-M.~W. Hwu, ``Profile-assisted instruction scheduling,'' \emph{International Journal of Parallel Programming}, 1994.

\bibitem{sla-400ms-2}
Y.~Chen, T.~Farley, and N.~Ye, ``Qos requirements of network applications on the internet,'' \emph{Information Knowledge Systems Management}, 2004.

\bibitem{z3-solver}
L.~De~Moura and N.~Bj{\o}rner, ``Z3: An efficient smt solver,'' in \emph{Tools and Algorithms for the Construction and Analysis of Systems: 14th International Conference (TACAS 2008)}.\hskip 1em plus 0.5em minus 0.4em\relax Springer, 2008, pp. 337--340.

\bibitem{dapple}
S.~Fan, Y.~Rong, C.~Meng, Z.~Cao, S.~Wang, Z.~Zheng, C.~Wu, G.~Long, J.~Yang, L.~Xia \emph{et~al.}, ``Dapple: A pipelined data parallel approach for training large models,'' in \emph{Proceedings of the 26th ACM SIGPLAN Symposium on Principles and Practice of Parallel Programming}, 2021, pp. 431--445.

\bibitem{swapadvisor}
C.-C. Huang, G.~Jin, and J.~Li, ``Swapadvisor: Pushing deep learning beyond the gpu memory limit via smart swapping,'' in \emph{Proceedings of the Twenty-Fifth International Conference on Architectural Support for Programming Languages and Operating Systems}, 2020, pp. 1341--1355.

\bibitem{gpipe}
Y.~Huang, Y.~Cheng, A.~Bapna, O.~Firat, D.~Chen, M.~Chen, H.~Lee, J.~Ngiam, Q.~V. Le, Y.~Wu \emph{et~al.}, ``Gpipe: Efficient training of giant neural networks using pipeline parallelism,'' in \emph{Advances in Neural Information Processing Systems}, 2019, pp. 103--112.

\bibitem{gems}
A.~Jain, A.~A. Awan, A.~M. Aljuhani, J.~M. Hashmi, Q.~G. Anthony, H.~Subramoni, D.~K. Panda, R.~Machiraju, and A.~Parwani, ``Gems: Gpu-enabled memory-aware model-parallelism system for distributed dnn training,'' in \emph{SC20: International Conference for High Performance Computing, Networking, Storage and Analysis}.\hskip 1em plus 0.5em minus 0.4em\relax IEEE, 2020, pp. 1--15.

\bibitem{flexflow}
Z.~Jia, M.~Zaharia, and A.~Aiken, ``Beyond data and model parallelism for deep neural networks,'' \emph{SysML 2019}, 2019.

\bibitem{toposort}
A.~B. Kahn, ``Topological sorting of large networks,'' \emph{Communications of the ACM}, 1962.

\bibitem{vilt}
W.~Kim, B.~Son, and I.~Kim, ``Vilt: Vision-and-language transformer without convolution or region supervision,'' in \emph{International Conference on Machine Learning}, 2021, pp. 5583--5594.

\bibitem{kim2012improving}
Y.~Kim, J.~Lee, T.~X. Mai, and Y.~Paek, ``Improving performance of nested loops on reconfigurable array processors,'' \emph{ACM Transactions on Architecture and Code Optimization (TACO)}, 2012.

\bibitem{megatron3}
V.~A. Korthikanti, J.~Casper, S.~Lym, L.~McAfee, M.~Andersch, M.~Shoeybi, and B.~Catanzaro, ``Reducing activation recomputation in large transformer models,'' \emph{Proceedings of Machine Learning and Systems}, 2023.

\bibitem{chimera}
S.~Li and T.~Hoefler, ``Chimera: efficiently training large-scale neural networks with bidirectional pipelines,'' in \emph{Proceedings of the International Conference for High Performance Computing, Networking, Storage and Analysis}, 2021, pp. 1--14.

\bibitem{superscaler}
Z.~Lin, Y.~Miao, G.~Liu, X.~Shi, Q.~Zhang, F.~Yang, S.~Maleki, Y.~Zhu, X.~Cao, C.~Li \emph{et~al.}, ``Superscaler: Supporting flexible dnn parallelization via a unified abstraction,'' \emph{arXiv preprint arXiv:2301.08984}, 2023.

\bibitem{swin-v2}
Z.~Liu, H.~Hu, Y.~Lin, Z.~Yao, Z.~Xie, Y.~Wei, J.~Ning, Y.~Cao, Z.~Zhang, L.~Dong \emph{et~al.}, ``Swin transformer v2: Scaling up capacity and resolution,'' in \emph{Proceedings of the IEEE/CVF conference on computer vision and pattern recognition}, 2022, pp. 12\,009--12\,019.

\bibitem{pipedream}
D.~Narayanan, A.~Harlap, A.~Phanishayee, V.~Seshadri, N.~R. Devanur, G.~R. Ganger, P.~B. Gibbons, and M.~Zaharia, ``Pipedream: generalized pipeline parallelism for dnn training,'' in \emph{Proceedings of the 27th ACM Symposium on Operating Systems Principles}, 2019, pp. 1--15.

\bibitem{megatron2}
D.~Narayanan, M.~Shoeybi, J.~Casper, P.~LeGresley, M.~Patwary, V.~Korthikanti, D.~Vainbrand, P.~Kashinkunti, J.~Bernauer, B.~Catanzaro \emph{et~al.}, ``Efficient large-scale language model training on gpu clusters using megatron-lm,'' in \emph{Proceedings of the International Conference for High Performance Computing, Networking, Storage and Analysis}, 2021, pp. 1--15.

\bibitem{gpt-4}
OpenAI, ``{GPT-4 Introduction},'' \url{https://openai.com/product/gpt-4}, [Online; accessed May-2023.

\bibitem{hetpipe}
J.~H. Park, G.~Yun, M.~Y. Chang, N.~T. Nguyen, S.~Lee, J.~Choi, S.~H. Noh, and Y.-r. Choi, ``Hetpipe: Enabling large dnn training on (whimpy) heterogeneous gpu clusters through integration of pipelined model parallelism and data parallelism,'' in \emph{2020 USENIX Annual Technical Conference (USENIX ATC 20)}, 2020, pp. 307--321.

\bibitem{PyTorch}
{PyTorch Team}, ``{PyTorch},'' \url{https://pytorch.org//}, [Online; accessed Mar.-2022.

\bibitem{gpt}
A.~Radford, K.~Narasimhan, T.~Salimans, and I.~Sutskever, ``Improving language understanding by generative pre-training,'' \emph{arXiv preprint arXiv:1704.01444}, 2018.

\bibitem{zero}
S.~Rajbhandari, J.~Rasley, O.~Ruwase, and Y.~He, ``Zero: Memory optimizations toward training trillion parameter models,'' in \emph{SC20: International Conference for High Performance Computing, Networking, Storage and Analysis}.\hskip 1em plus 0.5em minus 0.4em\relax IEEE, 2020, pp. 1--16.

\bibitem{deepspeed}
J.~Rasley, S.~Rajbhandari, O.~Ruwase, and Y.~He, ``Deepspeed: System optimizations enable training deep learning models with over 100 billion parameters,'' in \emph{Proceedings of the 26th ACM SIGKDD International Conference on Knowledge Discovery \& Data Mining}, 2020, pp. 3505--3506.

\bibitem{rau1992register}
B.~R. Rau, M.~Lee, P.~P. Tirumalai, and M.~S. Schlansker, ``Register allocation for software pipelined loops,'' in \emph{Proceedings of the ACM SIGPLAN 1992 conference on Programming language design and implementation}, 1992, pp. 283--299.

\bibitem{zero-offload}
J.~Ren, S.~Rajbhandari, R.~Y. Aminabadi, O.~Ruwase, S.~Yang, M.~Zhang, D.~Li, and Y.~He, ``Zero-offload: Democratizing billion-scale model training,'' in \emph{2021 USENIX Annual Technical Conference (USENIX ATC 21)}, 2021, pp. 551--564.

\bibitem{megatron1}
M.~Shoeybi, M.~Patwary, R.~Puri, P.~LeGresley, J.~Casper, and B.~Catanzaro, ``Megatron-lm: Training multi-billion parameter language models using gpu model parallelism,'' \emph{arXiv preprint arXiv:1909.08053}, 2019.

\bibitem{flava}
A.~Singh, R.~Hu, V.~Goswami, G.~Couairon, W.~Galuba, M.~Rohrbach, and D.~Kiela, ``Flava: A foundational language and vision alignment model,'' in \emph{Proceedings of the IEEE/CVF Conference on Computer Vision and Pattern Recognition}, 2022, pp. 15\,638--15\,650.

\bibitem{urpr}
B.~Su, S.~Ding, and J.~Xia, ``Urpr—an extension of urcr for software pipelining,'' in \emph{Proceedings of the 19th annual workshop on Microprogramming}, 1986, pp. 94--103.

\bibitem{piper}
J.~M. Tarnawski, D.~Narayanan, and A.~Phanishayee, ``Piper: Multidimensional planner for dnn parallelization,'' \emph{Advances in Neural Information Processing Systems}, 2021.

\bibitem{data-parallel}
P.~Team, ``{Distributed Data Parallelism},'' \url{https://pytorch.org/docs/stable/notes/ddp.html}, [Online; accessed Sep.-2022.

\bibitem{torchscript}
P.~Team, ``{TorchScript},'' \url{https://pytorch.org/docs/stable/jit.html}.

\bibitem{attention}
A.~Vaswani, N.~Shazeer, N.~Parmar, J.~Uszkoreit, L.~Jones, A.~N. Gomez, {\L}.~Kaiser, and I.~Polosukhin, ``Attention is all you need,'' \emph{Advances in neural information processing systems}, 2017.

\bibitem{branch-parallelism}
G.~Wang, X.~Fang, Z.~Wu, Y.~Liu, Y.~Xue, Y.~Xiang, D.~Yu, F.~Wang, and Y.~Ma, ``Helixfold: An efficient implementation of alphafold2 using paddlepaddle,'' \emph{arXiv preprint arXiv:2207.05477}, 2022.

\bibitem{tofu}
M.~Wang, C.-c. Huang, and J.~Li, ``Supporting very large models using automatic dataflow graph partitioning,'' in \emph{Proceedings of the Fourteenth EuroSys Conference 2019}, 2019, pp. 1--17.

\bibitem{simvlm}
Z.~Wang, J.~Yu, A.~W. Yu, Z.~Dai, Y.~Tsvetkov, and Y.~Cao, ``Simvlm: Simple visual language model pretraining with weak supervision,'' \emph{arXiv preprint arXiv:2108.10904}, 2021.

\bibitem{nphard}
G.~J. Woeginger, ``Exact algorithms for np-hard problems: A survey,'' in \emph{Combinatorial Optimization—Eureka, You Shrink!}, 2003, pp. 185--207.

\bibitem{sla-400ms}
X.~Xiao, \emph{Technical, commercial and regulatory challenges of QoS: An internet service model perspective}.\hskip 1em plus 0.5em minus 0.4em\relax Morgan Kaufmann, 2008.

\bibitem{gspmd}
Y.~Xu, H.~Lee, D.~Chen, B.~Hechtman, Y.~Huang, R.~Joshi, M.~Krikun, D.~Lepikhin, A.~Ly, M.~Maggioni \emph{et~al.}, ``Gspmd: General and scalable parallelization for ml computation graphs,'' \emph{arXiv preprint arXiv:2105.04663}, 2021.

\bibitem{mt5}
L.~Xue, N.~Constant, A.~Roberts, M.~Kale, R.~Al-Rfou, A.~Siddhant, A.~Barua, and C.~Raffel, ``mt5: A massively multilingual pre-trained text-to-text transformer,'' \emph{arXiv preprint arXiv:2010.11934}, 2020.

\bibitem{vocab500k}
B.~Zheng, L.~Dong, S.~Huang, S.~Singhal, W.~Che, T.~Liu, X.~Song, and F.~Wei, ``Allocating large vocabulary capacity for cross-lingual language model pre-training,'' in \emph{Empirical Methods in Natural Language Processing}, 2021, pp. 3203--–3215.

\bibitem{alpa}
L.~Zheng, Z.~Li, H.~Zhang, Y.~Zhuang, Z.~Chen, Y.~Huang, Y.~Wang, Y.~Xu, D.~Zhuo, E.~P. Xing \emph{et~al.}, ``Alpa: Automating inter-and $\{$Intra-Operator$\}$ parallelism for distributed deep learning,'' in \emph{16th USENIX Symposium on Operating Systems Design and Implementation (OSDI 22)}, 2022, pp. 559--578.

\bibitem{alpa-schedule}
Y.~Zhuang, L.~Zheng, Z.~Li, E.~Xing, Q.~Ho, J.~Gonzalez, I.~Stoica, H.~Zhang, and H.~Zhao, ``On optimizing the communication of model parallelism,'' \emph{Proceedings of Machine Learning and Systems}, 2023.

\end{thebibliography}
%%%%%%%%%%%%%%%%%%%%%%%%%%%%%%%%%%%%

\end{document}